\documentstyle[12pt,epsfig]{article}
\def\lsim{\raise0.3ex\hbox{$<$\kern-0.75em\raise-1.1ex\hbox{$\sim$}}}
\def\gsim{\raise0.3ex\hbox{$>$\kern-0.75em\raise-1.1ex\hbox{$\sim$}}}
\def\noi{\noindent}
\def\nn{\nonumber}
\def\bea{\begin{eqnarray}}  \def\eea{\end{eqnarray}}
\def\beq{\begin{equation}}   \def\eeq{\end{equation}}

\def\beeq{\begin{eqnarray}} \def\eeeq{\end{eqnarray}}

\begin{document}

\vbox to 2 truecm {}

\centerline{\bf BARYON STOPPING AND HYPERON ENHANCEMENT}
\vskip 3 truemm
\centerline{\bf IN THE IMPROVED DUAL PARTON MODEL}
\vspace{1 truecm}

\centerline{\bf A. Capella\footnote{e-mail : capella@qcd.th.u-psud.fr} and C. A. Salgado\footnote{e-mail :
salgado@qcd.th.u-psud.fr}} \vskip 0.3truecm
\author{\bf A. Capella\footnote{e-mail : capella@qcd.th.u-psud.fr} and C. A. Salgado\footnote{e-mail :
salgado@qcd.th.u-psud.fr}} 
\centerline{Laboratoire de Physique Th\'eorique
\footnote{Unit\'e Mixte de Recherche  (CNRS) UMR 8627}}
\centerline{Universit\'e de Paris XI, B\^atiment 210, F-91405 Orsay Cedex, France}
\vspace{1cm}

\begin{abstract}
We present an improved version of the dual parton model which contains a 
new realization of the
diquark breaking mechanism of baryon stopping. We reproduce in this way 
the net baryon yield in
nuclear collisions. The model, which also considers strings originating 
from diquark-antidiquark
pairs in the nucleon sea, reproduces the observed yields of $p$ and 
$\Lambda$ and their
antiparticles and underestimates cascades by less than 50 \%. However, 
$\Omega$'s are
underestimated by a factor five. Agreement with data is restored by 
final state
interaction, with an averaged cross-section as small as $\sigma = 0.14$ 
mb. Hyperon yields
increase significantly faster than antihyperons, in agreement with experiment.
\end{abstract}

\vskip 3 truecm
\noi LPT Orsay 99-20  \par
\noi March 1999

\newpage
\pagestyle{plain}
\baselineskip=24 pt

\section{Introduction} \hspace{\parindent} A striking feature of heavy ion collisions is the
huge stopping of the participating nucleons. At CERN energies, the deep minimum in the net
baryon rapidity distribution ($\Delta B = B - \bar{B}$) at $y^* \sim 0$, observed in $pp$
collisions, has been practically filled up in a central collision of heavy ions \cite{1r,2r}.
For central $Pb$-$Pb$ collisions, the value of this density is five times larger than the
corresponding value in $pp$ -- scaled by the average number of participants.
Note that the total number of $B - \bar{B}$ (i.e. integrated over rapidity), exactly satisfies
scaling in the number of participants, due to baryon number conservation. This shows the
dramatic change in the shape of the $B - \bar{B}$ (and $p - \bar{p}$) rapidity distributions
between $pp$ and central $Pb$-$Pb$ collisions. Such a change is usually referred to as baryon
stopping. \par

All independent string models of hadronic and nuclear collisions in their original form completely fail to
reproduce this important feature of heavy ion collisions. In the dual parton model (DPM)
\cite{3r} and in the quark gluon string model (QGSM) \cite{4r}, the do\-mi\-nant contribution to
particle production in $pp$ collisions at $\sqrt{s} \sim 20$~GeV, consists in two $qq$-$q$
strings, which produce, after fragmentation, two baryons in the fragmentation regions of the
colliding protons. Starting with the Lund model, which initially had a single string, the above
mechanism of particle production has been adopted in most current string models. In these
models, there is some amount of stopping due to energy conservation. This produces an increase
of the net baryon yield at mid rapidities between NN and central $Pb$-$Pb$ collisions, which is
typically of a factor 2 \cite{3r} -- more than two times smaller than the observed one. Hence, the
dramatic failure of all these models to reproduce the observed stopping. \par

Actually, a possibility to slow down the net baryon in $pp$ collisions, was introduced a long
time ago by Rossi and Veneziano \cite {5r}. In their approach, the baryon is viewed as three
valence quarks bound together by three strings each one with a quark at one end and with the
other end joining in a point called string junction. This string junction carries momentum as
well as the baryon quantum number. Rossi and Veneziano pointed out that the string junction
could migrate to mid rapidities with a distribution in $d\sigma /dx \sim 1/\sqrt{x}$ (or
$d\sigma /dy \sim \exp (- 1/2 |(y - y_{max}|)$. This corresponds to an annihilation
cross-section which decreases with energy like $s^{-1/2}$. In refs. \cite{6r} and \cite{7r} a
distribution of the string junction in $x^{-1}$ (i.e. flat in rapidity) was proposed,
corresponding to an annihilation cross-section which reaches a constant asymptotic value (of 1
to 2~mb). Here we adopt the first approach. However, we do not rule out the second possibility
-- which would have important consequences at the energies of the future heavy ion colliders
\cite {8r}. \par

The above stopping mechanism has been recently introduced in heavy ion collisions \cite{8r}
\cite{9r} and implemented in the Hijing \cite{10r} and Venus \cite{12r}
Monte Carlos. However, the introduction of the Rossi-Veneziano mechanism does not explain by
itself why the stopping is larger in central heavy ion collisions than in $pp$. In ref.
\cite{8r}, a mechanism to enhance stopping in heavy ion collisions was proposed. It was based
on the separation of the $pp$ cross-section $\sigma_{pp} = \sigma_{pp}^{DP} + \sigma_{pp}^{DB}$
into a diquark breaking ($DB$) and a diquark preserving ($DP$) piece, and on the assumption that
the diquark can be broken in any inelastic collision. These results in a $DB$ cross section in
$pA$ and $AA$ collisions which increases faster with $A$ than the $DP$ one. The drawback of this
approach is that it requires some fine tuning. The value of $\sigma_{pp}^{DB}$ has to be small
enough in order not to contradict the $pp$ and $pA$ data (where stopping is comparatively
small) and large enough to produce the large  stopping observed in central heavy ion collisions.
\par

In a recent publication \cite{13r}, a new formulation of the $DB$ mechanism has been introduced
in which this drawback is avoided (i.e. one can have $\sigma_{pp}^{DB}$ negligeably small at
CERN energies, and still have an important effect in central $Pb$-$Pb$ collisions). In the
present paper we use the formulation of \cite{13r} to compute the rapidity distributions of
$B - \bar{B}$ in hadronic and nuclear collisions. We obtain a reasonable agreement with
experiment\footnote{Note, however, that similar results are obtained using the approach of
ref. \protect{\cite{8r}} with the $DB$ mechanism of Fig.~2.}.  \par

Another striking feature of the CERN heavy ion program is the strong increase 
of the yields of
hyperons and anti-hyperons per participant between $pp$ or $pA$ and central $AB$ collisions. This
increase obeys to the hierarchy $\Omega > \Xi > \Lambda$ (i.e. the large number of strange
quarks in $Y$, the larger the increase) \cite{14r,15r}. In two recent publications \cite{13r}
\cite{16r}, it has been shown that a baryon stopping mechanism of the
type described  above produces a substantial increase of the hyperon yields according to this
hierarchy. The physical reason for this increase is quite obvious 
\cite{17r}. Since in the diquark breaking component the net baryon
is formed out of 
three sea quarks around the string junction, the probability of producing
hyperons is strongly enhanced -- especially for $\Omega$'s since its probability of
production in the conventional diquark fragmentation mechanism is zero. In the
present paper we extend the results of \cite{13r} in two directions. First, we study
the rapidity distributions of protons and hyperons  in
$pA$ and $AA$ collisions (in refs. \cite{13r} and \cite{16r} the analysis was
restricted to the rapidity window $|y^*| < 0.5$). Second, we show how the four free
parameters of ref. \cite{13r} can actually be reduced to two. This makes the model
more predictive, especially for the antihyperon over hyperon ratios. \par

While the yield of $\Lambda$'s and, to a large extent, of cascades can be described by the
model, that of omegas is underestimated by almost an order of magnitude. The same conclusion
has been reached in \cite{13r,16r}. We show that final state interaction, with an averaged
cross-section as small as $\sigma = 0.14$~mb \cite{13r}, allows to describe all 
hyperon and
antihyperon yields. \par

A similar value of $\sigma$ was found in ref. \cite{18r} in the hadron gas model. It was
argued there that, due to this small value of $\sigma$, interactions in a hadron gas could
not drive the system to chemical equilibrium (the process would be too slow). We find,
indeed, that the effect of final state interaction in $p$ and $\Lambda$ production is very
small. Its effect on $\Xi$ production is moderate. Only for such a rare process as $\Omega$
production its effect is very important -- making the $\Omega + \bar{\Omega}$ yield five
times larger than the value obtained without final state interaction. \par

Due to the diquark breaking component, we obtain an increase of hyperons substantially larger
than the one of antihyperons, i.e. the ratio between $Pb$-$Pb$ and $pPb$ yields is substantially
larger for $Y$ than for $\bar{Y}$. This effect is enhanced by final state interaction. It
has been observed experimentally \cite{15r}.\par

The paper is organized as follows. In Section 2 we describe the baryon stopping mechanism
and compute the net baryon ($B - \bar{B}$) rapidity distributions in $pp$, $pPb$, $SS$ and
$Pb$-$Pb$ collisions. In Section 3 we compute the rapidity distributions of 
$p-\bar{p}$,
and $Y-\bar{Y}$. In Section 4
we describe $B\bar{B}$ pair production from strings containing sea diquarks or
antidiquarks at one of their ends and show how the $A$-dependence of $\bar{B}$ production
is increased. In Section 5 we study the effect of the final state interaction, 
separately
on the $Y$ and $\bar{Y}$ yields. Section 6 contains a discussion of our  results.
Conclusions are given in Section 7.

\section{Baryon stopping}
\hspace{\parindent} A fragmentation string mechanism in which the $x \to 0$ and $x \to 1$
behaviour of the fragmentation functions is controlled by Regge intercepts has been introduced
in ref. \cite{19r}. In the case of net baryon production, it consists of a sum of two terms
as depicted in Fig.~1. In the original Lund fragmentation scheme \cite{20r} only the first term
is considered. The second one was introduced later -- the so-called pop-corn mechanism. Even
with the inclusion of the second component (Fig.~1b), this mechanism leads to the production of
too fast baryons and fails completely to reproduce the observed stopping in heavy ion
collisions. This fragmentation scheme (including the component in Fig.~1b) will be referred to
as the conventional or diquark preserving ($DP$) mechanism. Following ref. \cite{13r} we
introduce the baryon stopping mechanism showed in Fig.~2. It will be referred to as the diquark
breaking ($DB$) component. In this component, the rapidity distribution of the produced net
baryon $\Delta B = B - \bar{B}$ in a $N$-$N$ collision is

\beq
\label{1e}
{dN_{DB}^{\Delta B} \over dy} (y) = C_{n_1,n_2} \left [ Z_+^{1/2} \left ( 1 - Z_+
\right )^{n_1-3/2} + Z_-^{1/2} \left ( 1 - Z_- \right )^{n_2-3/2} \right ] \eeq

\noi where $Z_{\pm} = \exp (\pm y - y_{max})$, $n_1$ and $n_2$ are the average number of
collisions suffered by the two colliding nucleons and $C_{n_1,n_2}$ is determined from the
normalization to two. The factor $Z^{1/2}$ has already been discussed in the Introduction.
The factor $(1 - Z)$ gives the behavior near $y = y_{max}$. There is some uncertainty
concerning its power \cite{8r}. The value in eq. (\ref{1e}) is obtained as follows. From
Fig.~2 we see that in order to produce the baryon at $y \sim y_{max}$ it is necessary to
slow down three quarks. Assuming they behave as $1/\sqrt{x}$ at the energies under
consideration \cite{4r}, we obtain a power $1/2$ for the case of Fig. 2 -- which corresponds
to $n = 2$. In the general case of $n$ inelastic collisions we obtain the power $n - 3/2$
in eq. (\ref{1e}). \par

The corresponding distribution in $AA$ collisions is \cite{13r} 

\beq
\label{2e}
{dN^{AA\to \Delta B} \over dy} (y) = {\bar{n}_A \over \bar{n}} \left [ \bar{n}_A \left (
{dN_{DP}^{\Delta B} \over dy} (y) \right )_{\bar{n}/\bar{n}_A} + \left ( \bar{n} - \bar{n}_A
\right ) \left ( {dN_{DB}^{\Delta B} \over dy} (y) \right )_{\bar{n}/\bar{n}_A} \right ] \ . 
\eeq

\noi Here $\bar{n}_A$ and $\bar{n}$ are the average number of participants of nucleus $A$ and the
average number of collisions, respectively. $dN_{DP}/dy$ is given by the conventional, diquark
preserving, hadronization mechanism for which we use the results of ref. \cite{22r}, and
$dN_{DB}/dy$ is given by eq. (\ref{1e}). The integral over $y$ of both rapidity distributions
is equal to two (baryon number conservation). \par

Let us discuss the physical meaning of eq. (\ref{2e}) in $pp$ interactions
($\bar{n}_A = 1$). We see that in the case of a single inelastic collision ($\bar{n} = 1$), we
recover the conventional $DP$ mechanism. The underlying assumption is that, in this case, the
string junction follows the valence diquark and baryon production takes place in the
conventional way. (We do not exclude a small admixture of the $DB$ component in this case, but
experimental data do not require its presence). Consider next the case of two
inelastic collisions ($\bar{n} = 2$). Here the underlying assumption is that there is an equal
probability (1/2) for the net baryon to be produced in any of the two collisions. However, in
only one of them can the string junction follow the valence diquark and fragment in the
conventional ($DP$) way. In the other one, the string junction is free and baryon production
takes place according to the $DB$ mechanism. The latter is responsible of (most of) the observed
baryon stopping. The generalization to $\bar{n}$ inelastic collisions and to $AA$ interactions
(eq. (\ref{2e}))  is then straightforward. \par

It is quite remarkable that such a simple mechanism, with no free parameter (modulo the
uncertainty in the power of $1 - Z$ in eq. (\ref{1e}), discussed above) gives a good
description of the present data on the net baryon rapidity distribution. \par

Note that in the case of $pp$ interactions at $\sqrt{s} \sim 20$~GeV, the two string
component $(\bar{n} = 1)$ dominates, 
and, as discussed above, we recover the usual $DP$
results. With increasing energies, the components with $\bar{n} \not= 1$ become
increasingly important and baryon stopping will increase. Eq. (\ref{1e}) (with
$\bar{n}_A = 1$) not only gives definite predictions concerning this increase but,
moreover, leads to specific qualitative features. In particular stopping will strongly
depend on the charged particle multiplicity. A low multiplicity event sample selects low
values of $\bar{n}$ where stopping will be comparatively small, while, at large
multiplicities, stopping will be larger. Such a feature has been observed recently at
HERA \cite{23r} -- and discussed in ref. \cite{24r} in a different theoretical framework.
\par

We turn next to the generalization of eq. (\ref{2e}) to asymmetric interactions such as
$pA$. In this case one has

\bea
\label{3e}
{dN^{pA \to \Delta B} \over dy}(y) &=& {dN_{DP}^{qq^A-q_v^p} \over dy} (y) + (\bar{n} - 1)
{dN_{DP}^{qq^A - q_s^p} \over dy}(y) + \nn \\ &+&{1 \over \bar{n}} \left [ {dN_{DP}^{q_v^A -
qq^p} \over dy}(y) + (\bar{n}-1) {dN_{DB}^{q_v^A - qq_p} \over dy} (y) \right ] \ .
 \eea        

\noi Here $dN^{qq^A - q_v^p(q_s^p)}/dy$ denotes the rapidity distribution of a string stretched
between a diquark of one of the $\bar{n}$ wounded nucleons of $A$ and a valence (or sea) quark
of the proton. (It is computed in DPM as a convolution of momentum distribution functions
and fragmentation functions). Since each of the wounded nucleons suffers a single inelastic
collision, only the $DP$ component is involved -- with each diquark fragmenting in the nucleus
fragmentation region $(y^* < 0)$. The terms in the bracket correspond to the fragmentation of
the incoming proton. Since it suffers $\bar{n}$ inelastic collisions, we have the $DP$
hadronization mechanism (with probability $1/\bar{n}$) and the $DB$ one (with probability
$(\bar{n} - 1)/\bar{n}$). The latter is now given by the first term of eq. (\ref{1e}). All
rapidity distributions, integrated over $y$, are equal to one in this case. \par

The $B - \bar{B}$ rapidity distributions obtained from eq. (\ref{2e}) in central $SS$ and
$Pb$-$Pb$ collisions are shown in Fig.~3. Note that at mid-rapidities, these distributions are
dominated by the $DB$ component. Not only the latter is proportional to $\bar{n}-\bar{n}_A$
(eq. (\ref{2e})), but, moreover, $dN_{DB}/dy$ is larger than $dN_{DP}/dy$ at mid rapidities (see
Table 1). Nevertheless, the existence of the two huge maxima of the $DP$ component in the
fragmentation regions, still shows up in the $AA$ distribution. For a given system, the detailed
shape of the $B - \bar{B}$ rapidity distribution depends on the power of $(1 - Z)$ in eq.
(\ref{1e}). As discussed above there is some theoretical uncertainty in the value of this power.
However, the variation of the shape of this rapidity distribution from one system to another is
a characteristic feature of the model. As seen in Fig. 3, the minimum at mid-rapidities is
gradually filled up from $pp$ to central $Pb$-$Pb$ collisions and, therefore, it is more
pronounced in $SS$ than in $Pb$-$Pb$.

\section{Net hyperon enhancement} 
\hspace{\parindent} In the previous section we have shown that baryon stopping can be
described using a new formulation of the diquark breaking ($DB$) mechanism. In this case,
depicted in Fig.~2, the string junction is surrounded by three sea quarks to produce the net
baryon. Therefore, not only the net proton yield $\Delta p = p - \bar{p}$ will be strongly
enhanced from $pp$ to central $AA$ collisions, 
but also the net hyperon yield $\Delta Y = Y -
\bar{Y}$. This is especially so for $\Omega$'s, which cannot be
produced at all with the $DP$ mechanism of Figs.~1a and 1b\footnote{A new component consisting in
a diquark which contains sea quarks has been introduced in \protect{\cite{25r}}. However, this
diquark is assumed to have the same momentum distribution as a diquark made out of two valence
quarks and, hence, produces baryons mainly in the fragmentation regions.}. More precisely the
ratio between yields in central $AA$ and $pA$ (or $pp$) collisions will obey to the hierarchy~:
$\Delta \Omega > \Delta \Xi > \Delta \Lambda > \Delta p$. This is in agreement with the results
of the WA97 \cite{15r} and NA49 \cite{1r,14r} collaborations. \par

The relative yields of the different baryon species will be determined by the strangeness
suppression factor $S/L$ where $S$ is the probability associated to the strange quark and $L$
the one associated to the light quarks ($u$ or $d$). We consider two possibilities~: $S = 0.10$
and $L = (1 - S)/2 = 0.45$ ($S/L = 0.22$) and $S = 0.13$ and $L = 0.435$ ($S/L =
0.3$). With baryons produced out of three sea quarks (Fig.~2) it is easy to see that the relative
yields are

\beq
\label{4e}
I_3 = 4L^3 : 4L^3 : 12L^2S : 3LS^2 : 3LS^2 : S^3
\eeq

\noi for $p$, $n$, $\Lambda + \Sigma$, $\Xi^0$, $\Xi^-$ and $\Omega$, respectively. Moreover,
we take, $\Sigma^+ + \Sigma^- = 0.6 \Lambda$. This reduction in the number of
charged  
$\Sigma$'s is due 
to resonance decay ($\Sigma (1385)P_{13}$ decays into $\Lambda \pi$
with an $88 \pm 2\ \%$ fraction). \par

It is interesting that, in spite of the huge hyperon enhancement observed experimentally, the
factors (\ref{4e}) lead (both with $S = 0.10$ and $S = 0.13$) to an overestimation of hyperon
production in $pPb$ collisions especially for $\Xi$'s and $\Omega$'s. In central $Pb$-$Pb$
collisions, net hyperon production is also overestimated -- except for $\Omega$'s. In ref.
\cite{13r} the following explanation of this hyperon excess was proposed: 
at present
energies, 
it may happen that the net baryon is not formed out of three sea quarks as in
Fig.~2, but, due to phase space limitation, a valence quark, at one of the ends of the string
where the baryon is produced, is picked up together with two sea quarks. Obviously, in this case
the strangeness production rate is substantially reduced (in particular, $\Omega$ production
is not possible in this case). The relative yields $I_3$ in (\ref{4e}) are then
changed into 

\beq
\label{5e}
I_2 = 2L^2 : 2L^2 : 4LS : S^2/2 : S^2/2 : 0 \quad .
\eeq 

\noi In the following, we introduce a free parameter $\alpha$ ($0 < \alpha < 1$) which
determines the admixture of $I_3$ and $I_2$ given by (\ref{4e}) and (\ref{5e}). More
precisely, we will take the relative yields given by

\beq
\label{6e}
I = \alpha \ I_3 + (\alpha - 1) I_2 \quad .
\eeq

\noi The best description of the data is obtained with $\alpha = 0.23$ for $S = 0.13$ and
$\alpha = 0.5$ for $S = 0.1$. The results are similar in the two cases~; the 
sensitivity to
the value of the strangeness suppression factor $S/L$ turns out to be quite small. The results
for the rapidity distribution of the net yields of $p$ and $\Lambda$ in
central $Pb$-$Pb$ collisions are given in Figs.~4 and 5. Note that our rapidity
distribution for $\Lambda - \bar{\Lambda}$ (Fig.~5) is broader than the estimates of
the NA49 collaboration \cite{1r}. This, in turn, produces some discrepancies in the $p
- \bar{p}$ yield (Fig.~4). Final data on $\Lambda$ and $\bar{\Lambda}$ are needed in order to
clarify the situation~; we shall come back to this point in section 6. The net $\Lambda$
rapidity distribution in central $SS$ collisions is shown in Fig.~6. The corresponding results
for minimum bias $pA$ collisions are given in Figs.~7 and 8. It is seen that the normalization of
the experimental data is larger than the 
theoretical one especially for $p - \bar{p}$. Note,
however, that by integrating over $y$ the experimental distributions one realizes that their
normalization is larger than the number of wounded nucleus in $pAu$ (given by the Glauber
model) by more than a factor 2. As pointed out in \cite{2r} this excess may be due to recoil
nucleons which are not completely disentangled from the wounded ones. This point needs
clarification. \par

Note that, in our model, the relative yields $I_3$ (eq. (\ref{4e})) should apply at higher
energies when the phase space limitations are less important. With $S/L = 0.3$, they would give
a ratio $\Xi^+ + \Xi^0/\Lambda + \Sigma \sim 0.3$ which is in agreement with Fermilab
\cite{26r} and SPS collider data \cite{27r}.

\section{Antibaryon production} 
\hspace{\parindent} In string models, $B\bar{B}$ pair production takes place via
diquark-antidiquark pair production in the string fragmentation. It turns out that at present
CERN energies only strings of type $qq$-$q$ have large enough invariant mass to produce 
$B\bar{B}$ pairs. This gives rise to a scaling of $\bar{B}$ yields in the number of
participants. (The number of $qq$-$q$ strings is proportional to the number of wounded nucleons).
Experimentally, the observed increase is much faster -- closer to a scaling in the number of
collisions. In order to solve this problem it was proposed some time ago \cite{22r,28r,29r}
to consider the production of $B\bar{B}$ pairs from diquark-antidiquark pairs in the sea
of the participating nucleons\footnote{A different mechanism based on string
junction-antijunction exchange has been proposed recently \protect{\cite{16r}}.}. The rapidity
distribution of antibaryons in $AA$ collisions is then given by

\beq
\label{7e}
{dN^{AA \to \bar{B}} \over dy} (y) = \bar{n}_A \left ( {dN_{string}^{\bar{B}} \over dy} (y)
\right )_{\bar{n}/\bar{n}_A} + \left ( \bar{n} - \bar{n}_A \right ) \left ( {dN_{sea}^{\bar{B}}
\over dy} (y) \right )_{\bar{n}/\bar{n}_A} \ . \eeq    

\noi The first term represents the conventional pair production in the string breaking process.
As discussed above, it is proportional to the number of participants. The second term
corresponds to pair production from a string having a sea diquark or antidiquark at one of their
ends. In DPM, the total number of strings is proportional to $\bar{n}$. Since the number of
strings with a valence diquark at one of their ends is proportional to $\bar{n}_A$, the number
of strings with a sea diquark at one of their ends is proportional to $\bar{n} - \bar{n}_A$. Of
course, pulling a diquark-antidiquark pair out the nucleon sea is dynamically suppressed -- in
the same way as its production in the string breaking process is suppressed as compared to
$q$-$\bar{q}$ production. Thus, we expect that in each individual string, the production of
$B\bar{B}$ pairs in the two components (sea and string) in eq. (\ref{7e}) are comparable. In
practice, the normalization of the second component is treated as a free parameter. However, it
turns out that the sea component is always smaller than the string one, not only at $y^* = 0$
(see Table 1) but also after integration over rapidity. Note that the string
with sea diquarks have a smaller invariant mass.\par

For the string term we use the results of ref. \cite{22r}. The absolute normalization of this
term was determined from a fit of the $pp$ data. 
For the $y$-dependence of the sea term, we also use the
results of ref. \cite{22r}. As discussed above, its absolute normalization is a free parameter.
This parameter is the same for all species of baryons. More precisely, since the baryons and
antibaryons in the sea component are made out of three sea quarks or antiquarks, the relative
yields of the different baryon species is again given by eq. (\ref{4e})\footnote{Note that in
this case $\alpha = 1$, i.e. no admixture of the type discussed in connection with the $DB$
component is present here.}. (We neglect here the small differences in the rapidity shapes
induced by the different baryon masses). We are left in this way with a single free parameter
for this new sea component. Therefore, we have a total number of two free parameters, one in the
diquark breaking component and one in the sea component -- plus the value of the strangeness
suppression factor $S/L$ for which two values (0.22 and 0.3) have been considered. Of course,
the conventional components $DP$ and string in eqs. (\ref{2e}) and (\ref{7e}) contain
several free parameters. However, as discussed above, these parameters have been fixed in
ref. \cite{22r} from a fit of the $pp$ data and are not changed here
\footnote{For this 
reason, the change in the strange suppression parameter $S/L$ from 
0.22 to 0.3 only
applies to the new components $DB$ and sea in eqs. (\protect{\ref{2e}}) and
(\protect{\ref{7e}}) (see Table 1).}. The values at $y^*=0$ of the various
components for the different baryon species are given in Table 1.\par

The results for the rapidity distributions of the $\Lambda+\bar{\Lambda}$, $\Xi^- + \Xi^+$,
$\Xi^-$ and $\Omega+\bar{\Omega}$ yields in central $Pb$-$Pb$ collisions at 158 GeV are
given in Figs.~9-12. The $p$, $\bar{p}$, $Y$, and $\bar{Y}$ yields at $|y^*| < 0.5$ in minimum
bias $pPb$ collisions as well as at four different centralities in
$Pb$-$Pb$ collisions, are given in Fig.~13. The corresponding ratios 
$R_Y = \bar{Y}/Y$ at $|y^*|
< 0.5$ are given in Fig.~14. It should be noted that the values of $R_Y$ 
are not absolute predictions of our model. They can be changed by changing the 
normalization of
the sea component in eq. (\ref{7e}). However, the ratios $R_p : R_{\Lambda} : R_{\Xi} :
R_{\Omega}$ are a characteristic feature of the model. They show an increase with the number of
strange quarks in the baryon. In Fig.~15 we show the various baryon yields 
at $y^*=0$ divided to the
average number of participants, $2\bar{n}_A$, 
normalized to the same quantity in $pPb$. A
discussion of these results is given in Section 6, after introducing final state
interaction. 

\section{Final state interaction}
\hspace{\parindent} In an attempt to explain the strong enhancement of  the  $\Omega+\bar\Omega$
yield observed by the WA97 collaboration \cite{15r}, we are going to use our results
for the baryon densities as initial conditions in the gain and loss differential
equations which govern final state interactions \cite{18r,30r}

\beq
{dN_i\over d^4x}=\sum_{k,\ell }\sigma_{k\ell}\ \rho_k(x)\ \rho_{\ell}(x)-
\sum_{k}\sigma_{ik}\ \rho_i(x)\ \rho_k(x) \quad .
\label{eq7}
\eeq

The first term in the r.h.s. of (\ref{eq7}) describes the production of particles of type $i$
resulting from the interaction of particles $k$ and $\ell$ with space-time densities $\rho(x)$
and cross-sections $\sigma_{k\ell}$ (averaged over the momentum distribution of the interacting
particles).  The second term describes the loss of particles of type $i$ due to
its interaction with particles of type $k$. We use cylindrical space-time
variables and assume boost invariance (i.e. the densities $\rho(x)$ are taken 
to be independent of $y$). If we furthermore assume that the dilution in time
of  the densities is mainly due to longitudinal motion, i.e.~:

\beq
\rho_i(\tau,y,\vec s)=\rho_i(\tau,\vec s)\ {\tau_0\over\tau} \quad ,
\label{eq8}
\eeq

\noi
where $\tau=\sqrt{t^2-z^2}$ is the proper time and $\vec s$ the transverse 
coordinate, eqs. (\ref{eq7}) can be written as \cite{30r}

\beq
\tau{d\rho_i\over d\tau}=\sum_{k,l}\sigma_{k\ell} \ \rho_k\ \rho_{\ell} -
\sum_{k}\sigma_{ik}\ \rho_i\ \rho_k \ \ .
\label{eq9}
\eeq

\noi
Here $\rho_i(y,\vec s,\vec b)=dN_i/dy d\vec sd\vec b$. Thus, at
fixed impact parameter $\vec b$, we have to know the rapidity densities
per unit of transverse area $d\vec s$. Our eqs. (\ref{2e}) and (\ref{7e}) do
give these rapidity densities -- the dependence on $\vec s$ and $\vec b$
is contained in the geometrical factors $\bar n_A$ and $\bar n$, given by
the Glauber model. 
In the following, we use nuclear profiles obtained from Saxon-Woods
nuclear densities using the three-parameter Fermi distribution of ref. 
\cite{31r}.
For the pion densities we use the DPM results of ref. \cite{32r}
-- where explicit expressions as a function of $\bar n_A$ and $\bar n$
are given. 

Eqs. (\ref{eq9}) have to be integrated from initial time $\tau_0$ to 
freeze-out time $\tau$. These equations are invariant under the change
$\tau\to c \tau$. Therefore the result depends only on the ratio $\tau/\tau_0$.
Following refs. \cite{32r,33r}, we use the (inverse) proportionality between
$\tau$ and $\rho$ and put $\tau/\tau_0 = \rho (y, \vec{s}, \vec{b})/\rho_{f_0}$. Here $\rho (y,
\vec{s}, \vec{b})$ are the initial densities given by our expressions obtained in previous
sections and $\rho_{f_0}$ is the freeze-out density. For the latter, we take the charged density
per unit rapidity in a $pp$ collision, i.e. $\rho_{f_0} = [3/\pi R_p^2](dN^-/dy)_{y^*=0} =
1.15~{\rm fm}^{-2}$ \cite{32r,33r}. \par

We have now to specify the channels that have been taken into account in our calculation.
They are~:

\beq
\pi N \to K \Lambda \ , \quad \pi N \to K \Sigma \ , \quad \pi \Lambda \to K \Xi \ , \quad \pi
\Sigma \to K \Xi \ , \quad \pi \Xi \to K \Omega  \quad ,  
\label{10e}
\eeq

\noi and the corresponding reactions for antiparticles. To be more precise, of all possible
charge combinations in (\ref{10e}), some are of the type shown in Fig.~16a, 
with annihilation
of a light quark pair and production of an $s$-$\bar{s}$. 
They have all been taken into
account with the same cross-section $\sigma = 0.14$~mb. All other reactions in 
(\ref{10e}) are of
the type shown in Fig.~16b. They have three quark lines in the $t$-channel 
(baryon exchange). Their
average cross-section are smaller than the one of Fig.~16a and have been
neglected\footnote{The reactions we have kept are~: $\pi^+ + n \to K^+\Lambda$, $\pi^-p \to
K^0\Lambda$, $\pi^- + n \to K^0 \Sigma^-$, $\pi^+p \to K^+\Sigma^+$, $\pi^-\Lambda \to
K^0\Xi^-$, $\pi^+\Lambda \to K^+\Xi^0$, $\pi^+\Sigma^- \to K^+\Xi^-$,
$\pi^-\Sigma^+ \to K^0\Xi^0$, $\pi^-\Xi^0 \to
K^0\Omega$ and $\pi^+ \Xi^- \to K^+\Omega$ for the reactions initiated by $\pi^+$ or $\pi^-$.
For all of them, as well as for the corresponding ones with antiparticles, we take $\sigma =
0.14$~mb. The reactions initiated by $\pi^0$ are either of the type of Fig.~16a or Fig.~16b,
depending on whether the $u\bar{u}$ or $d\bar{d}$ component of the $\pi^0$ is considered. For
this reason all these reactions have been included with cross-section $\sigma/2$.}. We have
also neglected all strangeness exchange reactions (Fig. 16c) 
$KN \longleftrightarrow \pi \Lambda$ etc. Although the
corresponding cross-sections are larger at threshold, this is no longer the case for the
cross-sections averaged over the momentum distributions of the interacting particles . (This
is due to their steep decrease with increasing energy (see \cite{18r}). Channels (\ref{10e})
are thus dominant due to the relations $\rho_N > \rho_{\Lambda} > \rho_{\Xi} > \rho_{\Omega}$
and $\rho_{\pi} > \rho_K$ between particle densities. The results, obtained after solving
numerically eqs. (\ref{eq9}), with our initial densities and a 
common value of the averaged
cross-section $\sigma = 0.14$~mb for all channels, are shown 
in our figures by a full line in the case $S/L=0.3$ and $\alpha=0.23$ and by a 
dashed-dotted line in the case $S/L=0.22$ and $\alpha=0.5$.
(A
comparable value of $\sigma$ has been obtained in ref. \cite{18r} in a hadron gas model). 

The
effect of the final state interaction is negligeably small in $pA$ collisions. In central $SS$
collisions its effect on the $p$ and $\Lambda$ yields is very small (less than 5~$\%$). The
effect increases with the number of strange quarks in the produced hyperon. In central $Pb$-$Pb$
collisions, with our value of the cross-section, it turns out to be comparatively small for $p$
and $\Lambda$ yields. However, it increases the $\Xi$ yields by up to 50~$\%$ and the $\Omega+\bar\Omega$
yield by a factor 5. Agreement with the WA97 data \cite{15r} is obtained in this way (Fig. 13).
\par

It is important to note that, due to the small value of $\sigma$, the final state
interaction has an important effect only on very rare processes such as $\Omega$ production.
It cannot drive the system into chemical equilibrium -- even locally.

\section{Discussion}
\hspace{\parindent} We discuss here the main features of our results. 
For $\Lambda +
\bar{\Lambda}$ our results for $Pb$-$Pb$ are slightly higher than the WA97 
data 
and grossely
underestimate NA49 ones at mid-rapidities (Fig.~9). Note, however, that the latter are very
preliminary and are currently under reanalysis. For central $SS$ collisions, where the NA35
data are final, we slightly underestimate their 
net $\Lambda$ yield and slightly overestimate
the total $\Lambda$ yield form NA36 \cite{34r}
(Fig.~6). However, the NA35 value for the $\bar{\Lambda}$ yield at mid
rapidities $0.75 \pm 0.15$ is about two times larger than our result. Note that this
experimental point looks ``anomalous'': compared with the WA97 value for the most central
rapidity bin in $Pb$-$Pb$ collisions, ($1.8 \pm 0.2$), there is an increase by a factor 2.4
-- whereas the number of participants increases by a factor 7. Note also that NA35 finds a
ratio $\bar{\Lambda}/\bar{p} = 1.9 \pm 0.7$ at mid-rapidities, while in our model this ratio
is always smaller than one (see Fig. 13). This important point needs clarification. In
particular, final values of this ratio in $Pb$-$Pb$ collisions are needed. \par

Concerning, the cascade yields, the NA49 data are published \cite{14r}. They are $30 \div 40$~\%
higher than the WA97 ones \cite{15r} at mid rapidities. Our results, after final state
interaction, are intermediate between the two sets of data -- but somewhat closer to the NA49
results (Figs.~10 and 11). The $\Omega + \bar{\Omega}$ yields are in agreement with the WA97
data, after final state interaction (Fig.~12 and 13). \par

As discussed in Section 4, in our model, the ratios $R_Y = \bar{Y}/Y$ increase with the number
of strange quarks in the baryon (Fig.~14). This tendency is also seen in the data. However, the
ratio of ratios $R_{\Xi}/R_{\Lambda}$ is somewhat too small in our model as compared to the
WA97 data \cite{15r}, but agrees with the NA49 ones \cite{1r,14r}. (Remember,
however, that the value of
$R_{\Lambda}$ from NA49 is preliminary). \par

Another characteristic feature of our approach is that, at mid-rapidities, hy\-pe\-rons are
more strongly enhanced than antihyperons.
As a consequence,
the ratio $\bar{Y}/Y$ decreases between $pPb$ and central $Pb$-$Pb$ collisions (Fig.~14). This is
due to the strong effect of the $DB$ component in the net baryon yield. Final state interaction
works in the same direction. This important feature of our results is  seen in 
the 
data \cite{15r}. \par

Finally, the WA97 collaboration has found that the increase of the hyperon and antihyperon
yields per participant (Fig.~15) increases faster than the number of participants between
$pPb$ and the first centrality bin in $Pb$-$Pb$. However, between the first and last
centrality bin all yields approximately scale with the number of participants. We find an
increase which is faster in the first case than in the second one. However, some mild increase
is left in $Pb$-$Pb$ (Fig.~15). 

\section{Conclusions} \hspace{\parindent} The large baryon stopping observed in central
heavy ion collisions at CERN e\-ner\-gy is not reproduced by any of the available independent string models
-- at least in their original form. We have modified the DPM by introducing a new
realization of the diquark breaking mechanism. We reproduce in this way the observed net
baryon yield. This mechanism also produces an important enhancement of net hyperons.
At this level, the new version of DPM presented here (which has also diquark-antiquark pairs in
the nucleon sea) remains strictly an independent string model. It reproduces with two free
parameters, the observed yields of $p$ and $\Lambda$ and their antiparticles in $pA$ and
$Pb$-$Pb$ collisions. Cascades in central $Pb$-$Pb$ collisions are underestimated
by less than 50 \% while $\Omega$'s are too small by a factor 5. Agreement
with experiment is restored by introducing final state interaction with an averaged cross-section
as small as $\sigma = 0.14$~mb. In this way, we depart from string
independence. However, with this small value of the cross-section, there is no
significant effect on the bulk of particle production. A comparable value of the
cross-section for final state interaction was obtained in ref. 
\cite{18r} from the
experimental data on the energy dependence of cross-sections, 
averaged over the momentum
distribution of the interacting particles obtained in the hadron gas model. The
smallness of this averaged cross-section lead the authors of \cite{18r} to argue that
strangeness phase space saturation would be too slow in a hadron gas. It is 
interesting that such a small value of the averaged cross-section allows to
reproduce the observed enhancement of multi-strange hyperons and antihyperons in central
$Pb$-$Pb$ collisions. \par 

The main features of our results are the following~: 1) The hyperon yields per participant
increase faster than antihyperon ones. As a consequence, 
the ratio $R_Y = \bar{Y}/Y$ decreases between
$pPb$ and central $Pb$-$Pb$ collisions~; 2) The ratios $R_Y$ increase with the number of
strange quarks in the hyperon~; 3) The increase of the $Y$ and $\bar{Y}$ yields per
participant is faster between $pPb$ and the first centrality bin in $Pb$-$Pb$ collisions
and slows down between the first and last centrality bins of WA97. All these features are also
present in the data.\\

\noi \subsection*{Acknowledgments}
\hspace{\parindent} It is a pleasure to thank N. Armesto, 
A. Casado, E. G. Ferreiro, A. B.
Kaidalov, C. Pajares
and J. Tran Thanh Van for discussions. We also thank R. Lietava, P.
Seyboth and O. Villalobos Baillie for information on the data.             
A.C. acknowledges partial support from a NATO grant 
OUTR.LG 971390. C.A.S. thanks Fundaci\'on
Caixa Galicia  from Spain for financial support.

\newpage
\section*{Figure Captions}
\vskip 1 truecm
\begin{description}
\item{\bf Fig. 1 :} Conventional diquark preserving (DP) fragmentation 
mechanism for net baryon production.

\item{\bf Fig. 2 :}  Example of diquark breaking (DB) diagram for net baryon production 
in $pA$ with two inelastic collisions.

\item{\bf Fig. 3 :} Rapidity distribution of the net baryon number ($B-\bar B$)
in central $SS$ (200 AGev/c) and $PbPb$ (158 AGev/c) collisions. The full lines
are obtained from eq. (2). The data are from ref [1,2] (for central $SS$ 
collisions the data are obtained as $B-\bar B=2\ (p-\bar p)+1.6\ 
(\Lambda -\bar \Lambda)$). Open circles
are data reflected about $y^*=0$ (errors not shown). The dotted line is
the result obtained without the DB component in the case of central $PbPb$ 
collisions. Due to baryon number conservation, these results are not affected
by final state interactions.

\item{\bf Fig. 4 :} Rapidity distributions for net proton production 
($p-\bar p$) in central $PbPb$ collisions at 158 AGev/c compared to the
results of ref. [1]. Open circles are data reflected about $y^*=0$ 
(errors not shown). The dashed line is our result without final state 
interactions with a strangeness suppression factor $S/L=0.3$, 
and the full line is the
corresponding result with final state interactions. The dotted-dashed
line corresponds to a suppression factor $S/L=0.22$ and with final state 
interactions. The dotted line is our result without DB component and
without final state interactions.

\item{\bf Fig. 5 :} Same as Fig. 4 for $\Lambda -\bar\Lambda$. Now the three
dotted lines are estimates by the NA49 Collaboration [1]. The experimental 
point at $y^*=0$ is form the WA97 Collaboration [14].

\item{\bf Fig. 6 :} Same as Fig. 5 for central $SS$ collisions. The data are
from NA35 [2] (circles). Also shown, the value of the total $\Lambda$ yield
measured by NA36 [32] (squares). The theoretical curves are computed with:
$\alpha=$ 0.23, $S/L=$ 0.3 (full line); 
$\alpha=$ 1, $S/L=$ 0.3 (dashed line); 
$\alpha=$ 0.23, $S/L=$ 0.22 (dotted line) and
$\alpha=$ 1, $S/L=$ 0.22 (dashed-dotted line)

\item{\bf Fig. 7 :} Same as Fig. 4 for minimum bias $pAu$ interactions. The
experimental data are from ref. [2].

\item{\bf Fig. 8 :} Same as Fig. 5 for minimum bias $pAu$ interactions. The
experimental data are from ref. [2].

\item{\bf Fig. 9 :} Same as Fig. 4 for $\Lambda +\bar\Lambda$. The 
point at
$y^*=0$ (circle) is from the WA97 Coll. [14]. The
squares are from NA49 [1].

\item{\bf Fig. 10 :} Same as Fig. 9 for $\Xi^-+\bar\Xi^+$. The NA49 data are 
from ref [13]. 

\item{\bf Fig. 11 :} Same as Fig. 9 for $\Xi^-$. The NA49 data are from ref 
[13]. 

\item{\bf Fig. 12 :} Same as Fig. 9 for $\Omega +\bar\Omega$. The experimental point
is from WA97 [14].

\item{\bf Fig. 13 :} Yields of $p$, $\Lambda$, $\Xi^-$, $\Omega+\bar\Omega$,
$\bar p$, $\bar\Lambda$ and $\bar\Xi^+$ for minimum bias $pPb$ (158 Gev/c) 
and central $PbPb$ collisions (158 AGeV/c) in four centrality bins. 
Experimental data are
 from WA97 [14] (black points) and NA49 [13] (open square). 
Full (dashed) lines are our results with 
(without) final state interactions for 
strangeness
suppression factor $S/L=0.3$. The dashed-dotted lines are our results with
final state interactions for
$S/L=0.22$.

\item{\bf Fig. 14 :} Ratios $\bar B/B$ at $y^*=0$ for minimum bias $pPb$ and 
$PbPb$ collisions in four different centrality bins at 158 AGev/c. Black
circles,
squares and triangle correspond to experimental data of WA97 [14] for $\bar
\Lambda/\Lambda$, $\bar\Xi^+/\Xi^-$ and $\bar\Omega/\Omega$ respectively. 
Open squares are
NA49 data [13]. Full lines are our results with final state interactions and
dashed lines without final state interactions, both for $S/L=0.3$.

\item{\bf Fig. 15 :} Same as Fig. 14 for the baryon yields 
at $y^*=0$ in $PbPb$ divided by the number of
participant nucleons relative to the same ratio in minimum bias $pPb$. 

\item{\bf Fig. 16a :} Quark diagrams for reactions (11) with light quark 
pair annihilation and $s-\bar s$ quark creation.

\item{\bf Fig. 16b :} Quark diagram for reactions (11) with three quark 
exchange in the t-channel 

\item{\bf Fig. 16c :} Quark diagram for strangeness exchange reactions.

\end{description}

\newpage
\section*{Table}
\vskip 1 truecm
\begin{center}
\begin{tabular}{|l|c|c|c|c|}
\hline
& & & & \\
&$p$ &$\Lambda$ &$\Xi^-$ &$\Omega$ \\
& & & & \\
\hline
& & & & \\
$dN_{sea}^{\bar{B}}/dy$ &$4.80\times 10^{-3}$ & $2.15\times 10^{-3}$ &$ 3.21\times 10^{-4}$& $3.20\times 10^{-5}$ \\ 
$dN_{DB}^{\Delta B}/dy$ &$9.12\times 10^{-2}$ & $3.00\times 10^{-2}$ & $2.86\times 10^{-3}$ & $1.23\times 10^{-4}$ \\ 
& & & & \\
\hline
& & & & \\
$dN_{DP}^{\Delta B}/dy$ &$6.90\times 10^{-2}$ & $1.40\times 10^{-2}$ & 0 & 0 \\
$dN_{string}^{\bar{B}}/dy$ &$8.50\times 10^{-3}$ & $2.26\times 10^{-3}$ & $1.65\times 10^{-4}$ & $5.07\times 10^{-6}$ \\ 
& & & & \\
\hline
& & & & \\
$dN_{sea}^{\bar{B}}/dy$ &$6.54\times 10^{-3}$ & $2.43\times 10^{-3}$& $2.98\times 10^{-4}$ & $2.4\times 10^{-5}$ \\ 
$dN_{DB}^{\Delta B}/dy$ &$9.56\times 10^{-2}$ & $2.62\times 10^{-2}$ & $2.29\times 10^{-3}$ & $1.23\times 10^{-4}$ \\ 
& & & & \\
\hline
\end{tabular}
\end{center}
\vskip 1 truecm
\noi {\bf Table 1 :} Values of the rapidity densities at $y^*=0$ in eqs. (\ref{2e}) and (\ref{7e})
for central $PbPb$ collisions ($\bar n_A=$ 178, $\bar n=$ 858) with
$\alpha=0.23$ and $S/L=0.3$ 
(first four lines) and with $\alpha=0.5$ and $S/L=0.22$
(last two lines). The DP and $string$ contributions are the same
in both cases. The value of the DP component for $\Xi^-$ is not exactly 0,
due to the fragmentation mechanism 
of Fig. 1b. However its value is very small as
compared to the other components and has been neglected.

\newpage

\centerline{\bf Figure 1}

\begin{center}
\epsfig{file=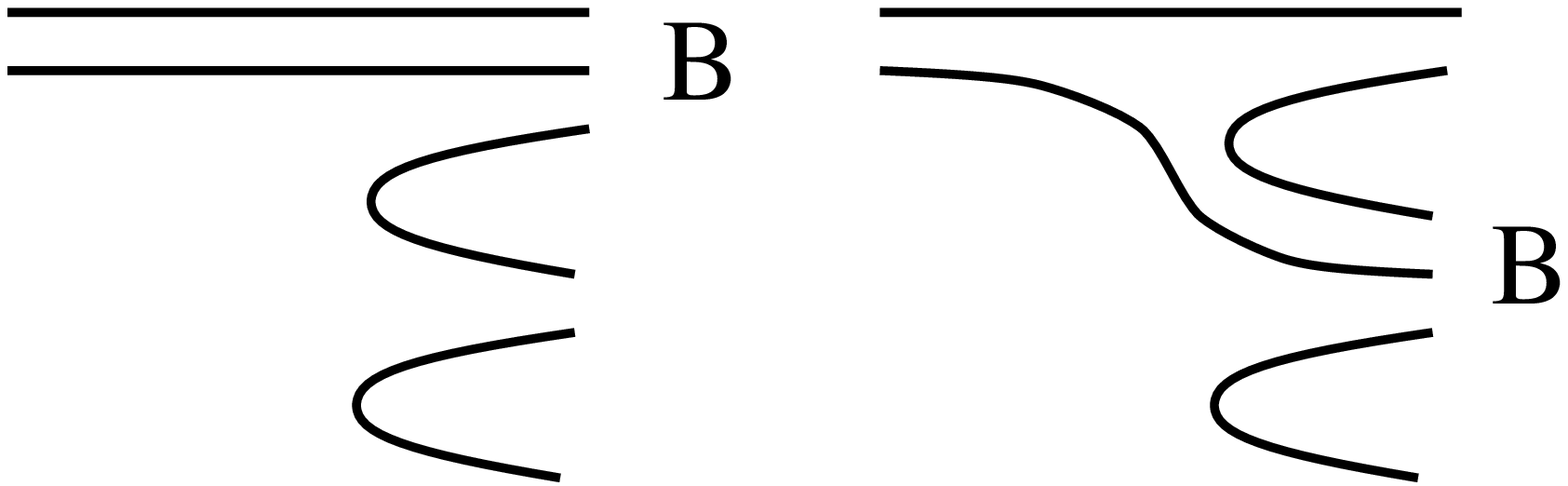,width=10.cm}
\end{center}

\vspace{1cm}
\centerline{\bf Figure 2}

\begin{center}
\epsfig{file=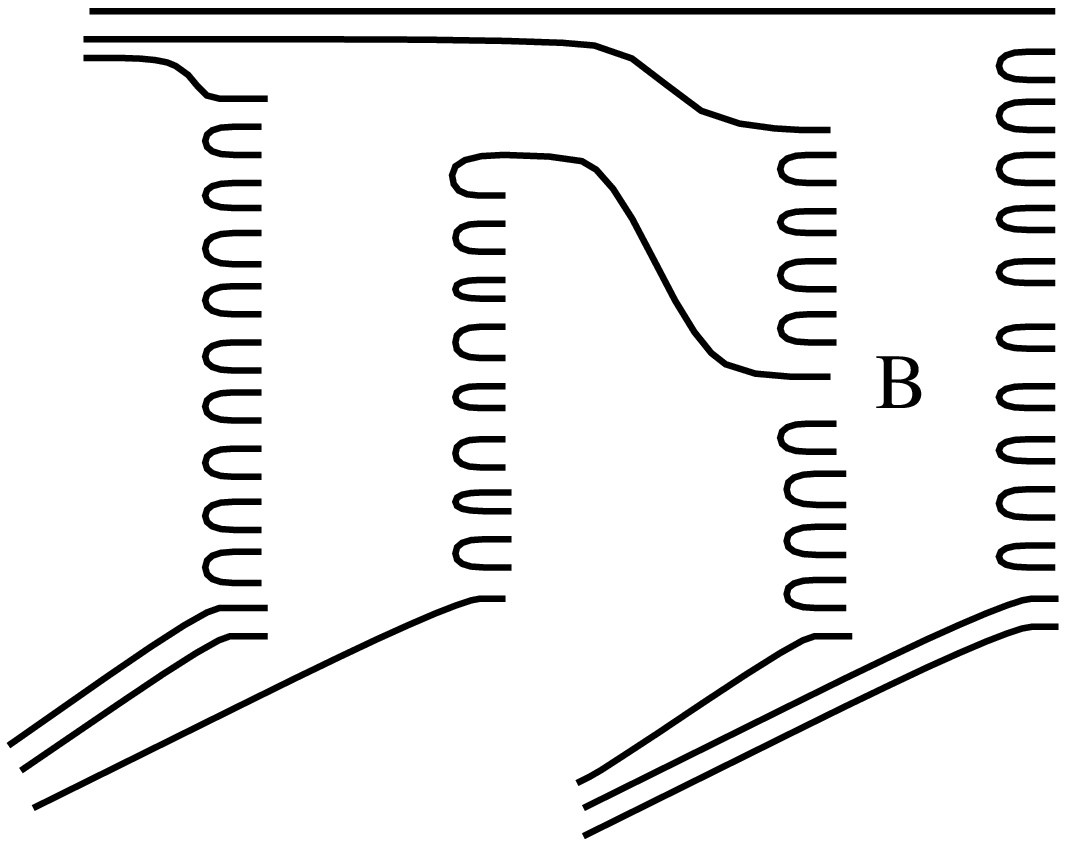,width=10.cm}
\end{center}

\newpage
\centerline{\bf Figure 3}

\begin{center}
\epsfig{file=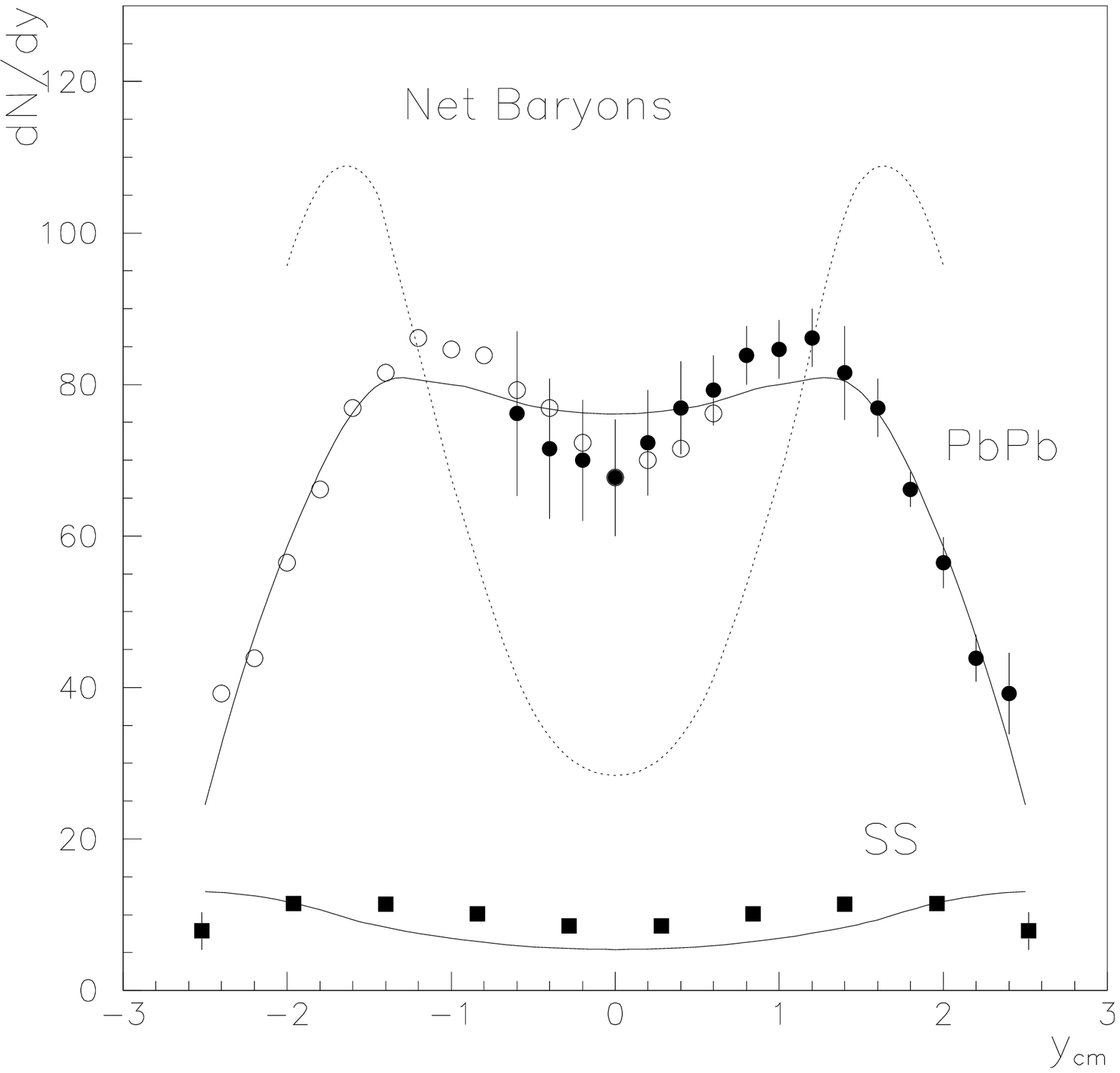,width=13.cm}
\end{center}

\newpage
\centerline{\bf Figure 4}

\begin{center}
\epsfig{file=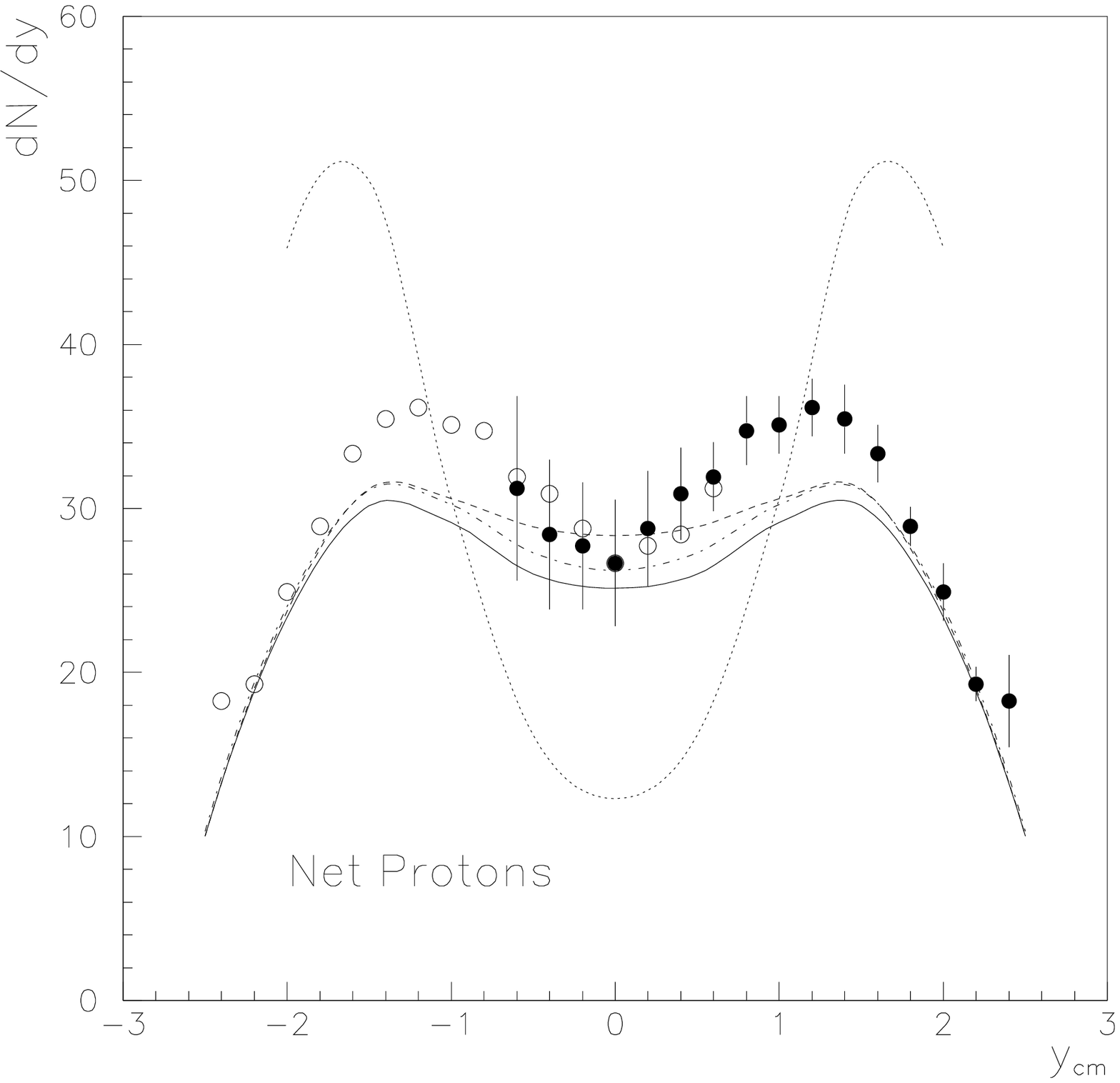,width=13.cm}
\end{center}

\newpage
\centerline{\bf Figure 5}

\begin{center}
\epsfig{file=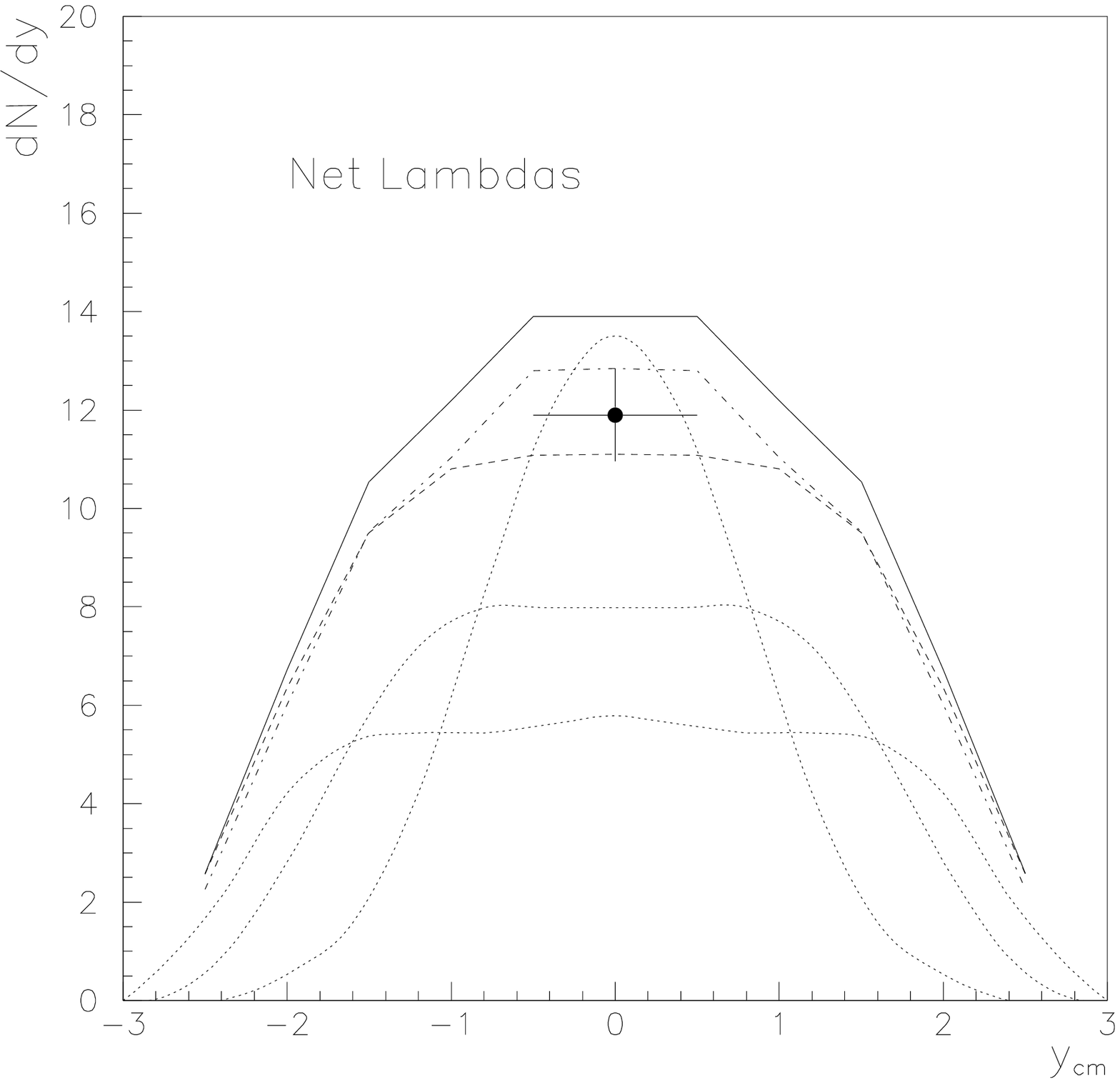,width=13.cm}
\end{center}

\newpage
\centerline{\bf Figure 6}

\begin{center}
\epsfig{file=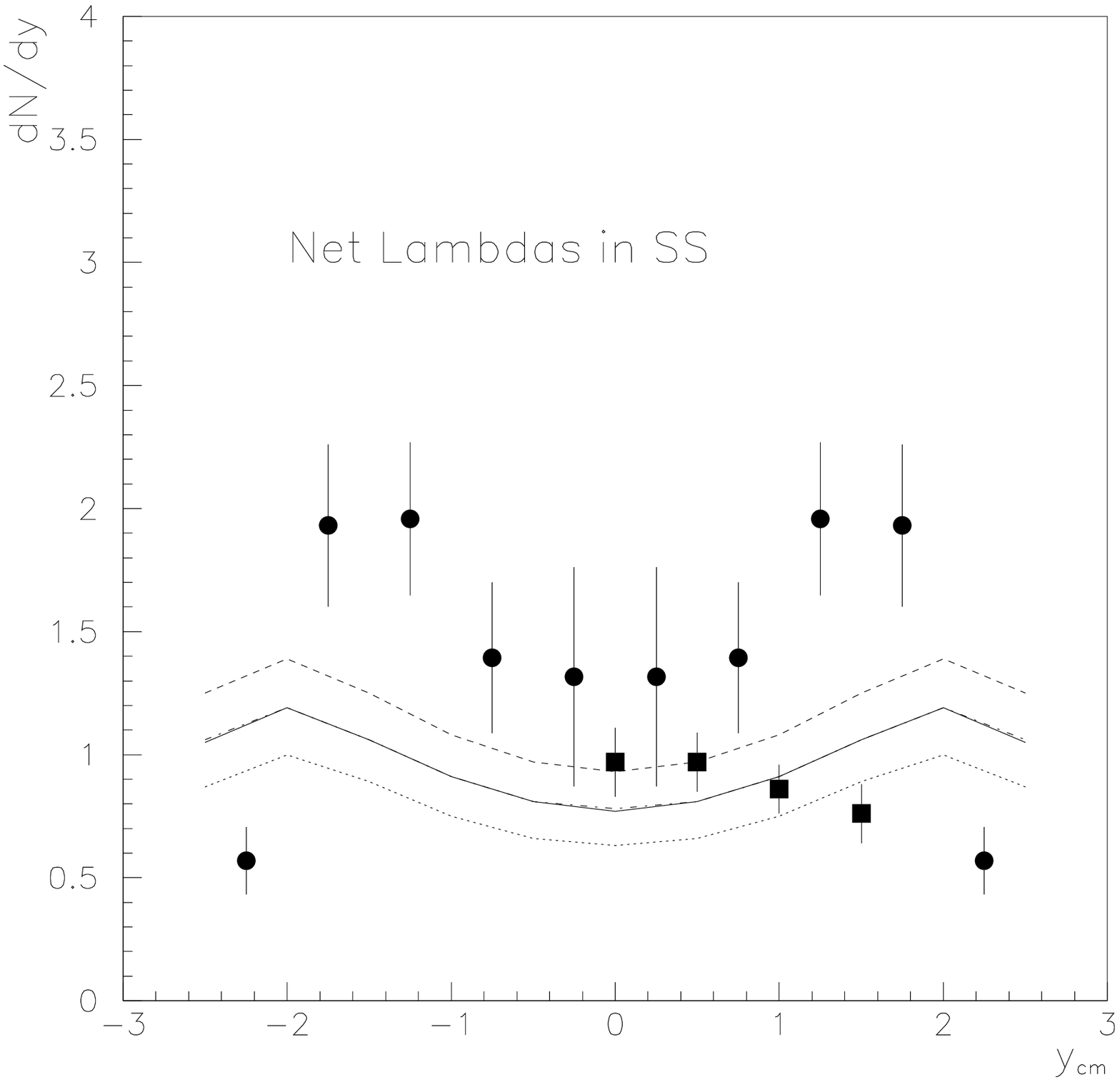,width=13.cm}
\end{center}

\newpage
\centerline{\bf Figure 7}

\begin{center}
\epsfig{file=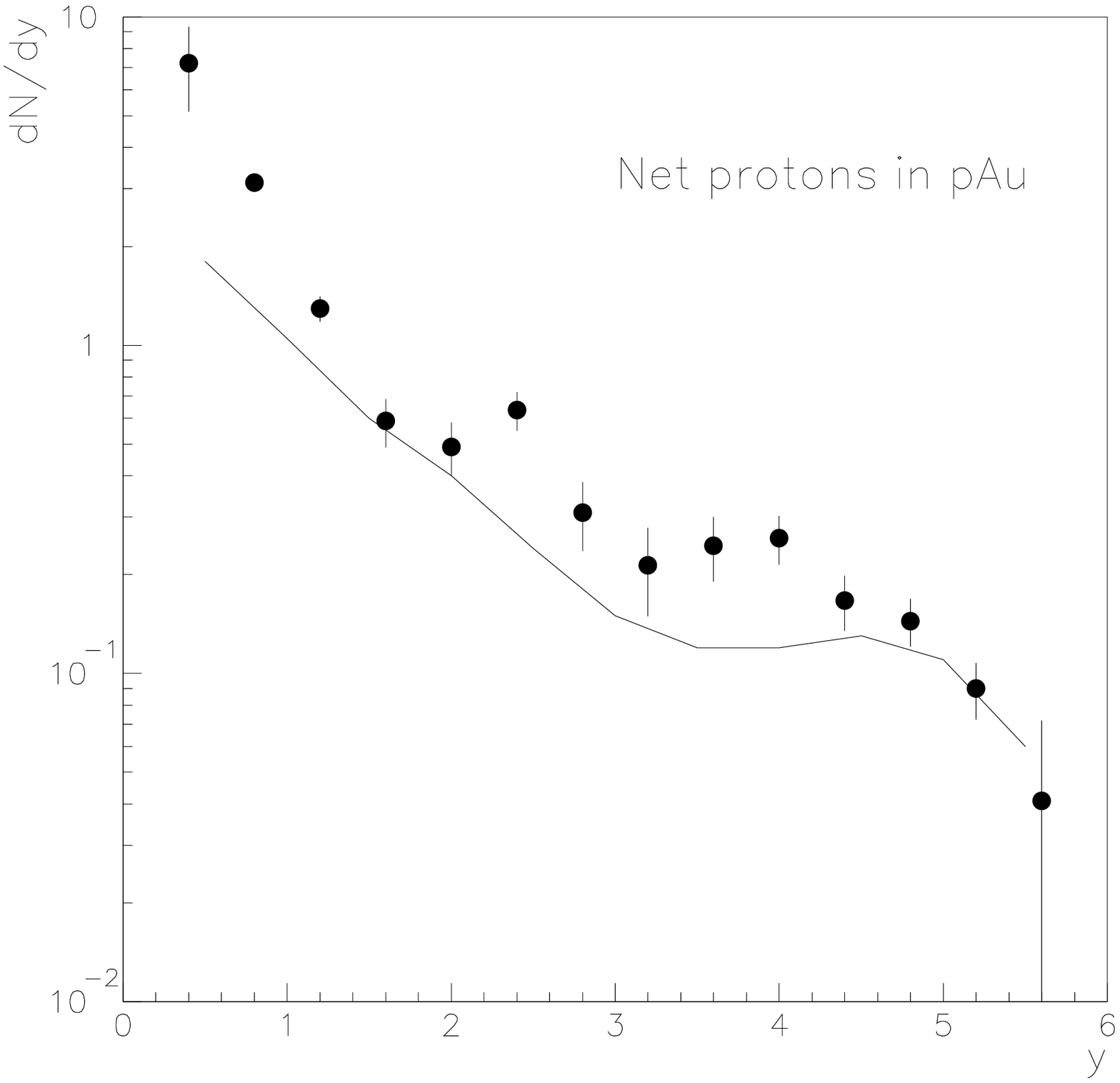,width=13.cm}
\end{center}

\newpage
\centerline{\bf Figure 8}

\begin{center}
\epsfig{file=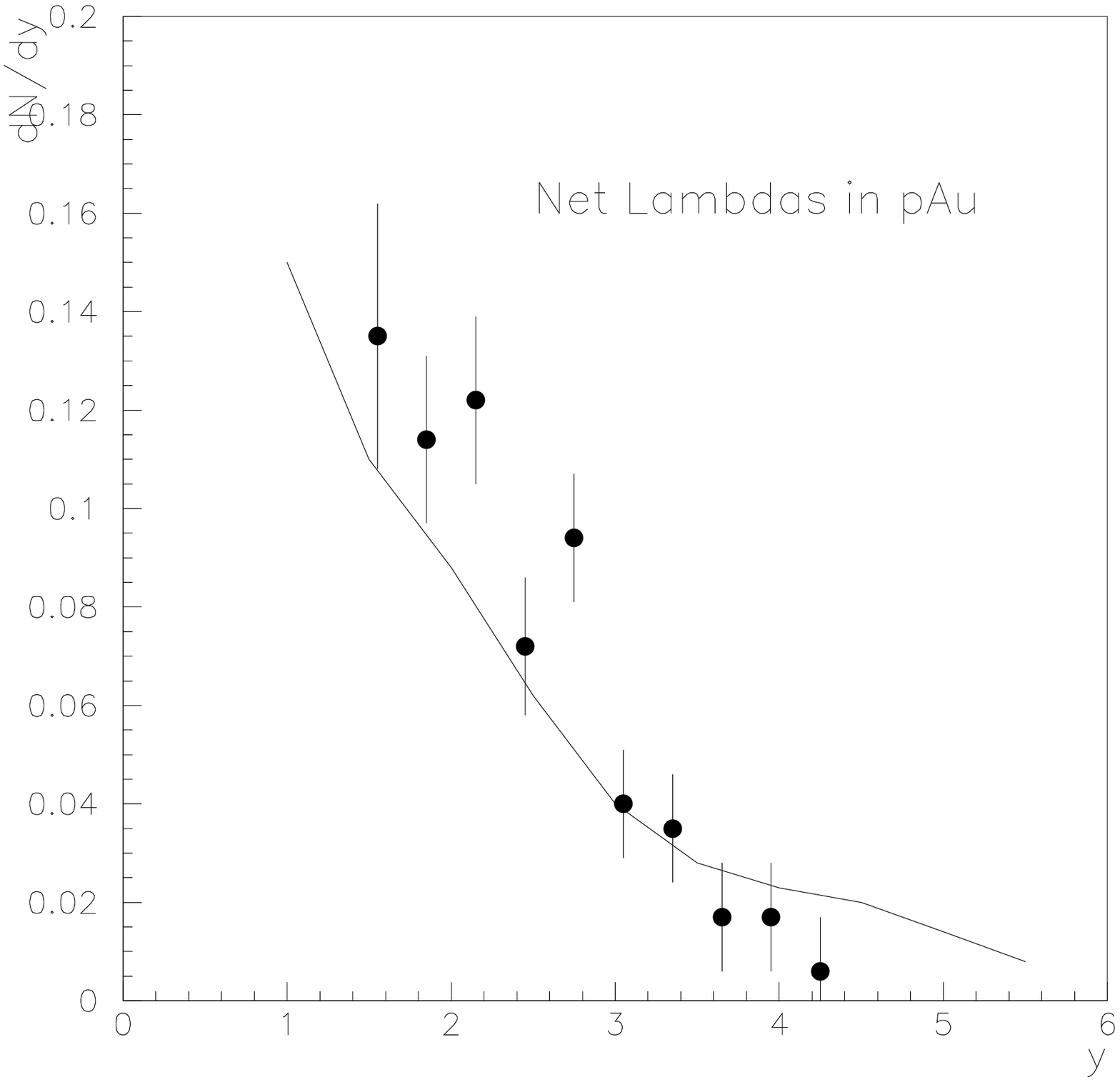,width=13.cm}
\end{center}

\newpage
\centerline{\bf Figure 9}

\begin{center}
\epsfig{file=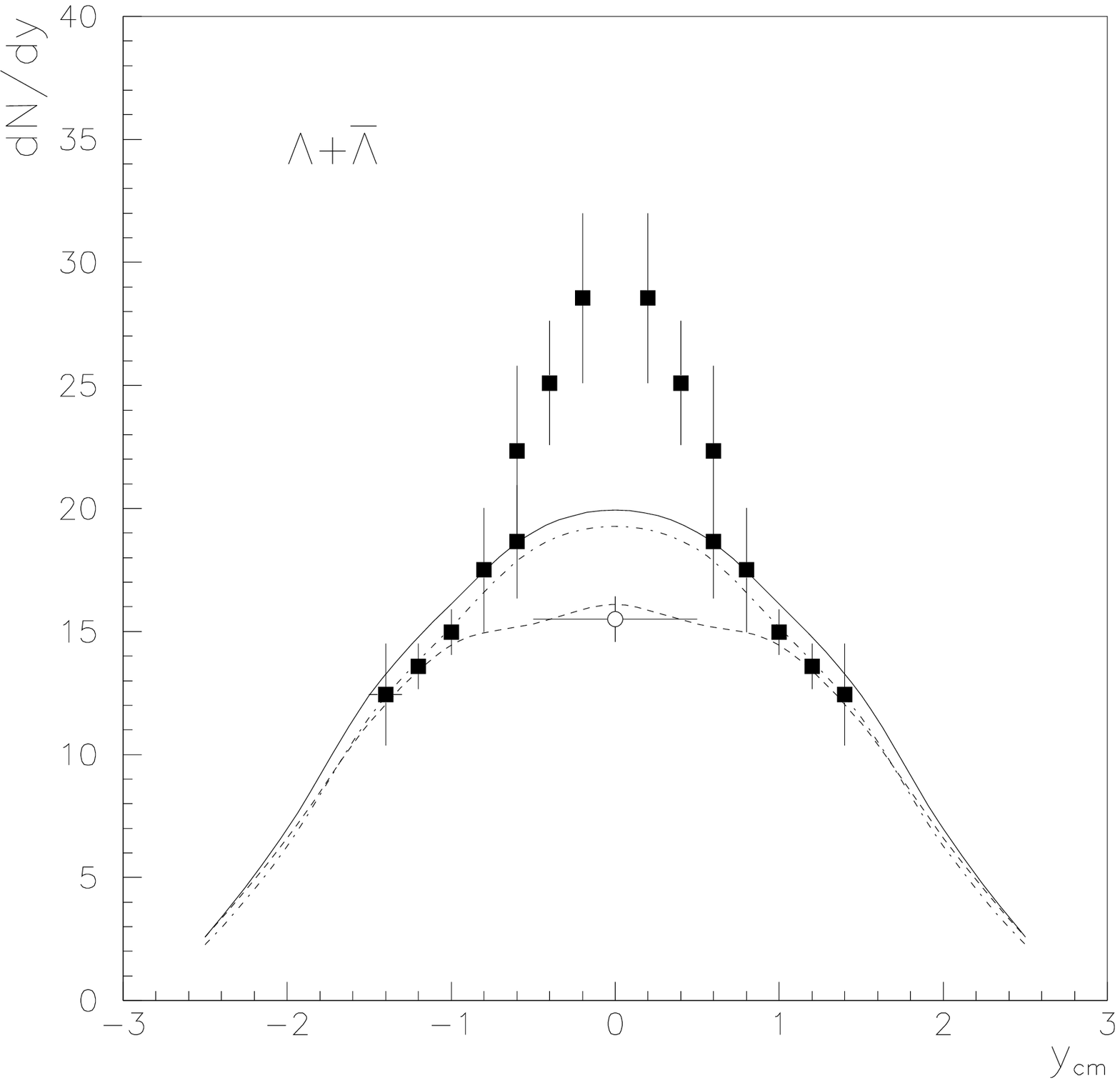,width=13.cm}
\end{center}

\newpage
\centerline{\bf Figure 10}

\begin{center}
\epsfig{file=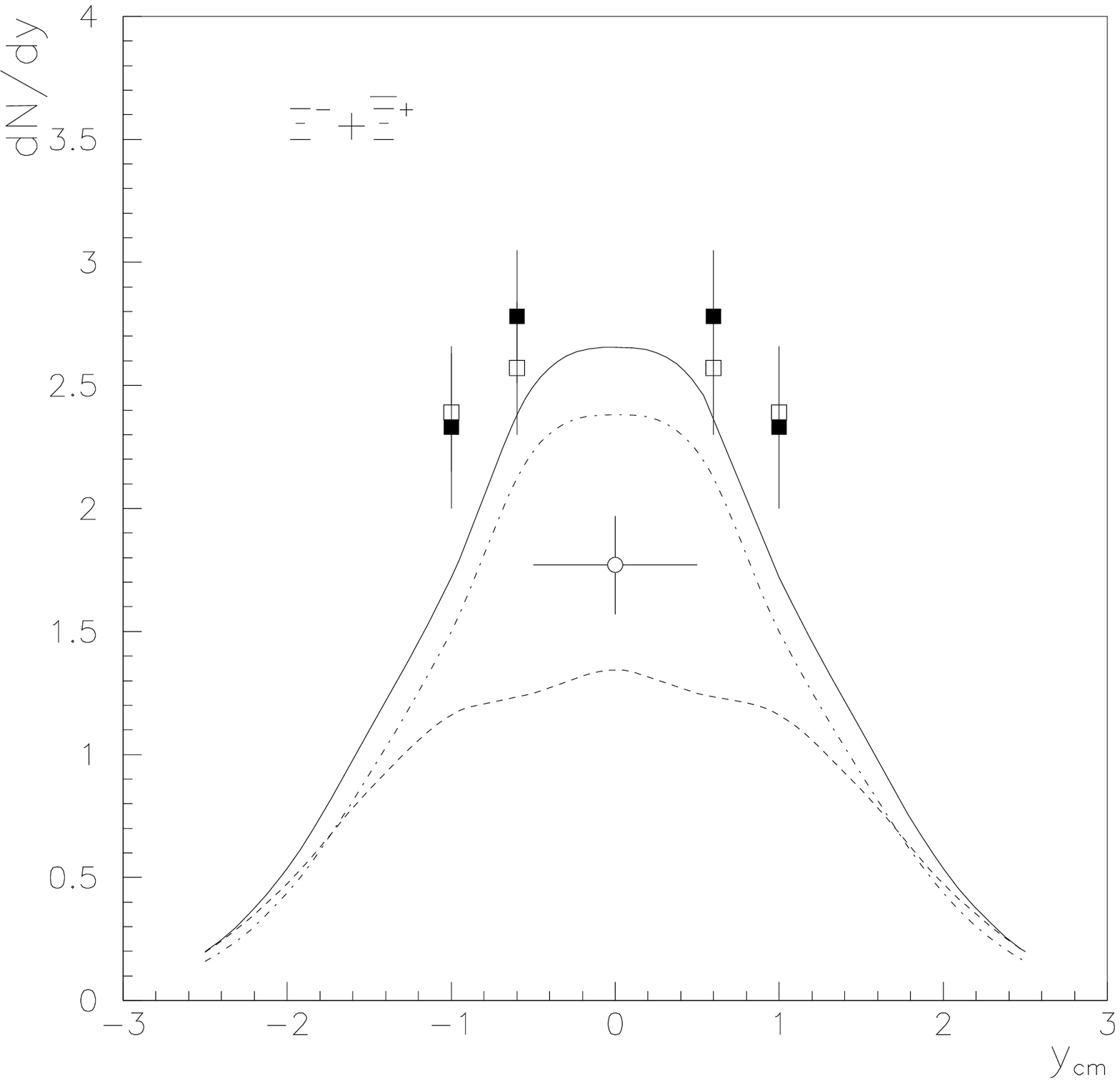,width=13.cm}
\end{center}

\newpage
\centerline{\bf Figure 11}

\begin{center}
\epsfig{file=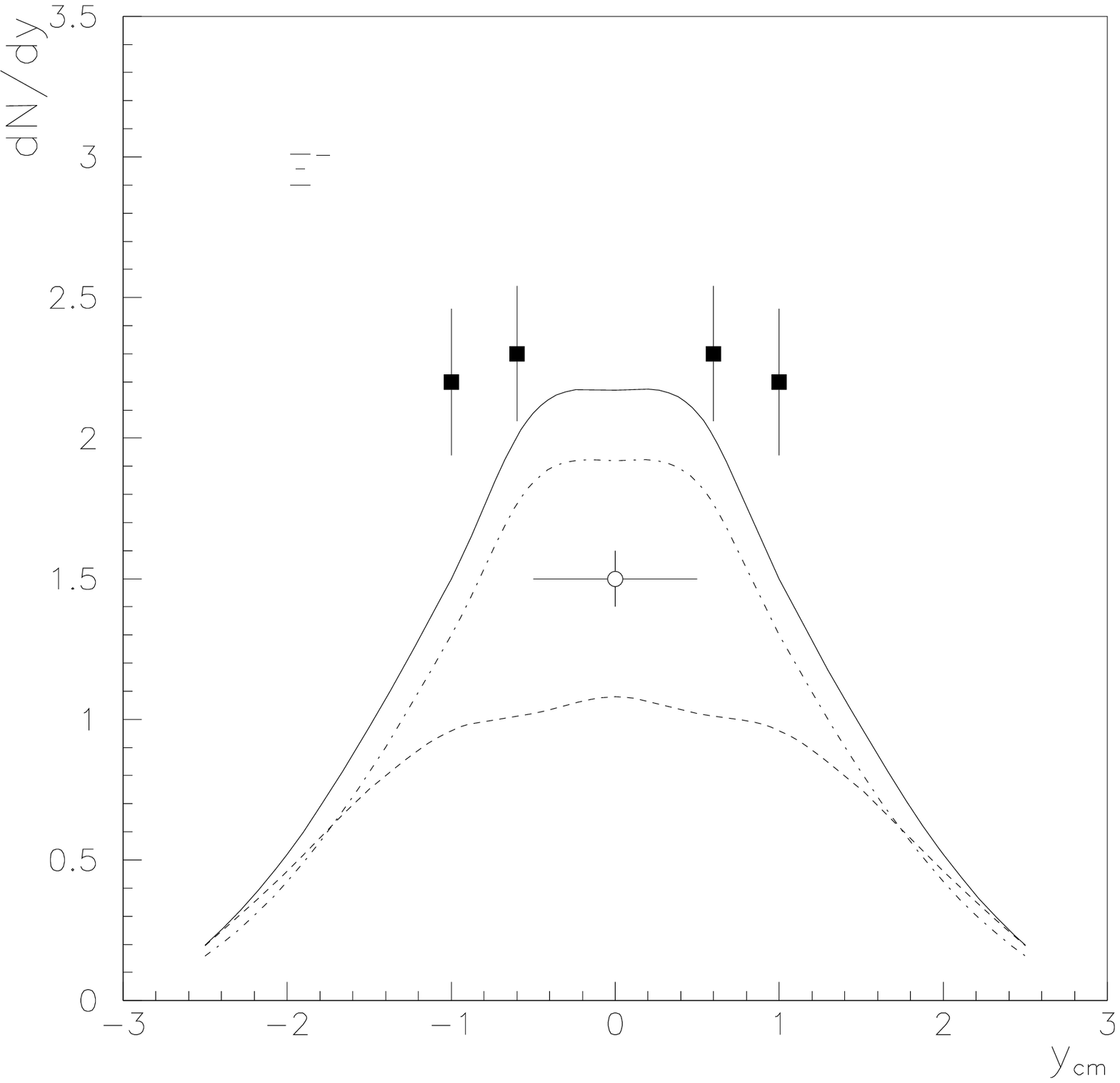,width=13.cm}
\end{center}

\newpage
\centerline{\bf Figure 12}

\begin{center}
\epsfig{file=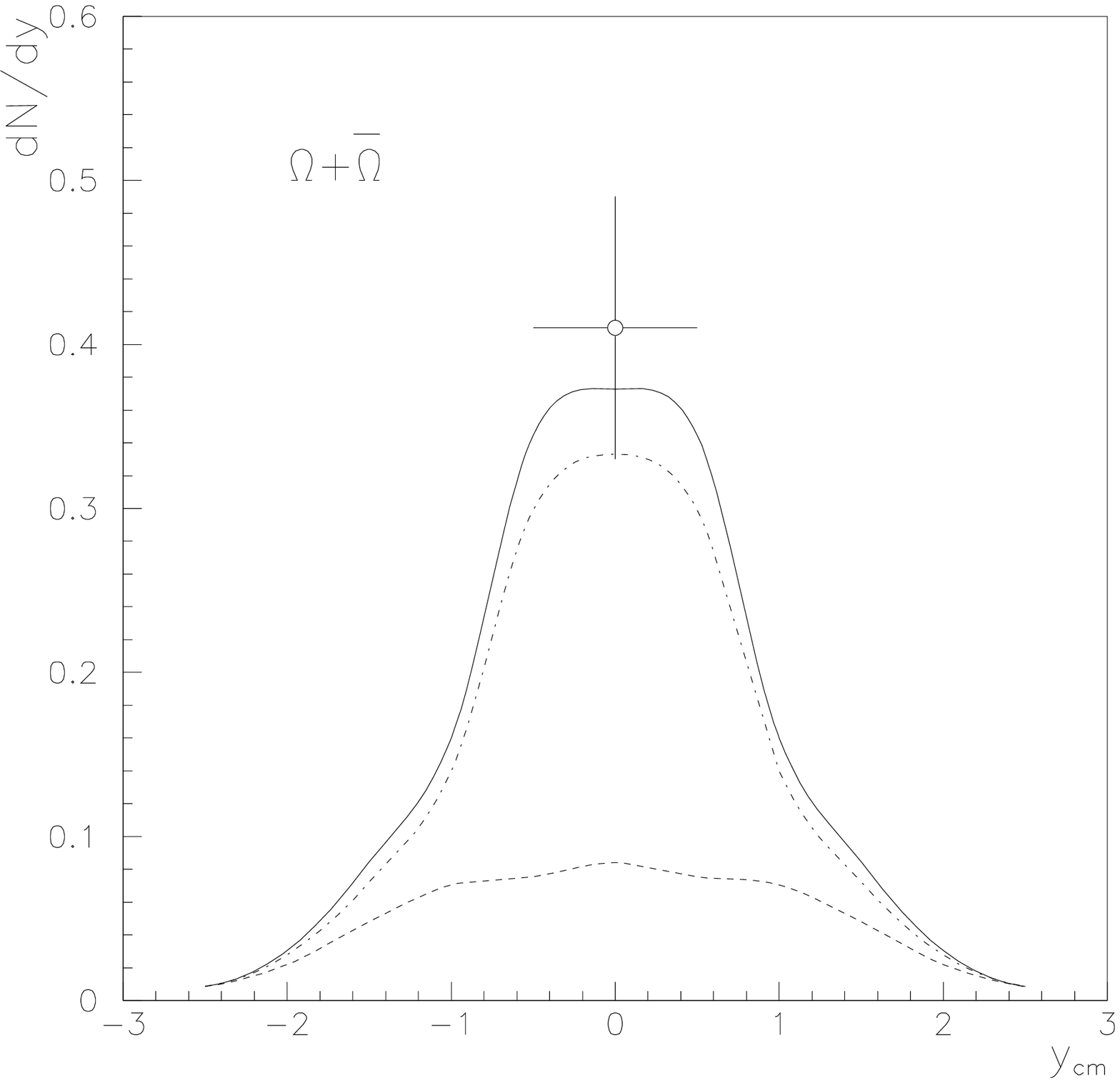,width=13.cm}
\end{center}

\newpage
\centerline{\bf Figure 13}

\begin{center}
\epsfig{file=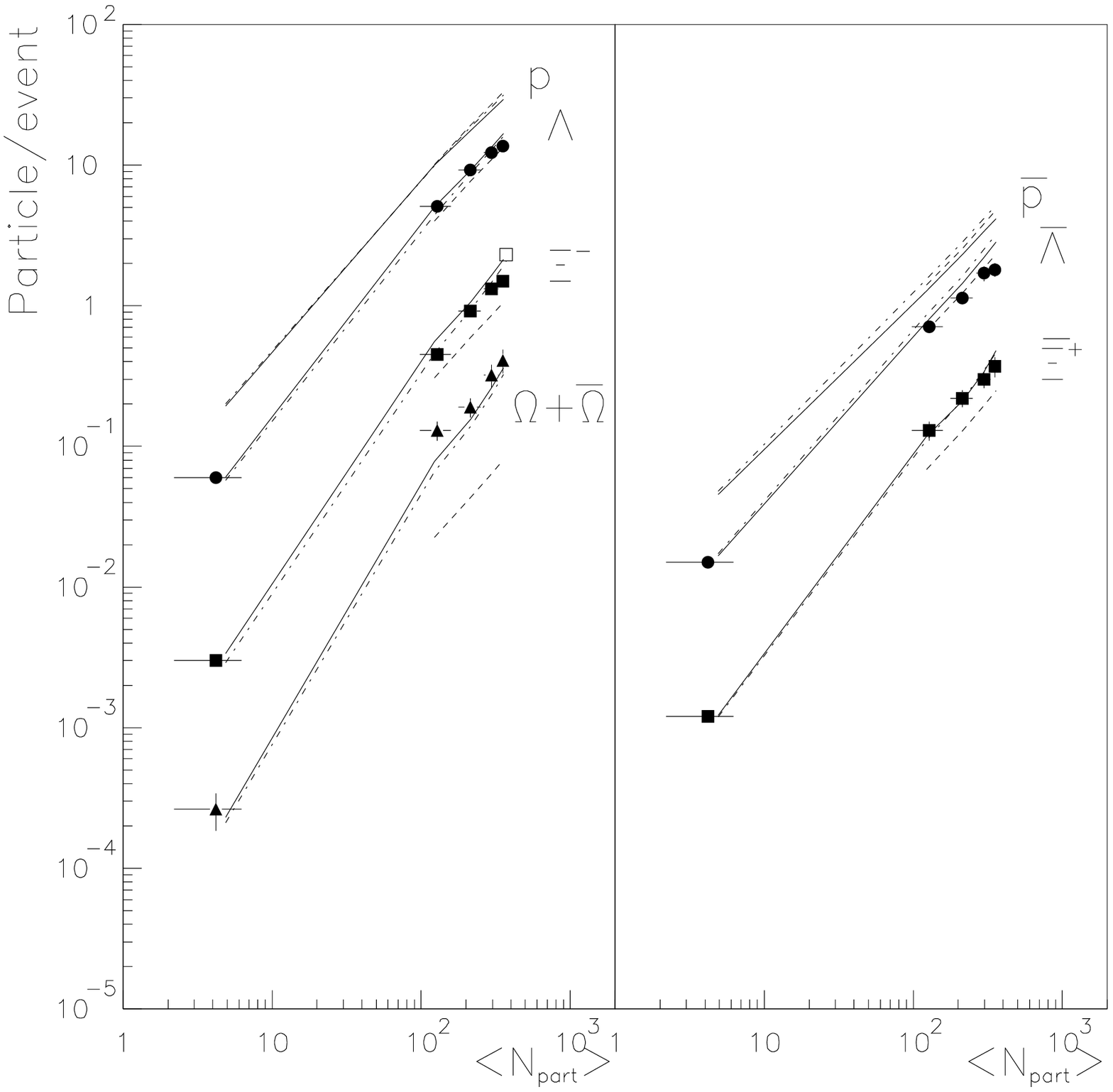,width=13.cm}
\end{center}

\newpage
\centerline{\bf Figure 14}

\begin{center}
\epsfig{file=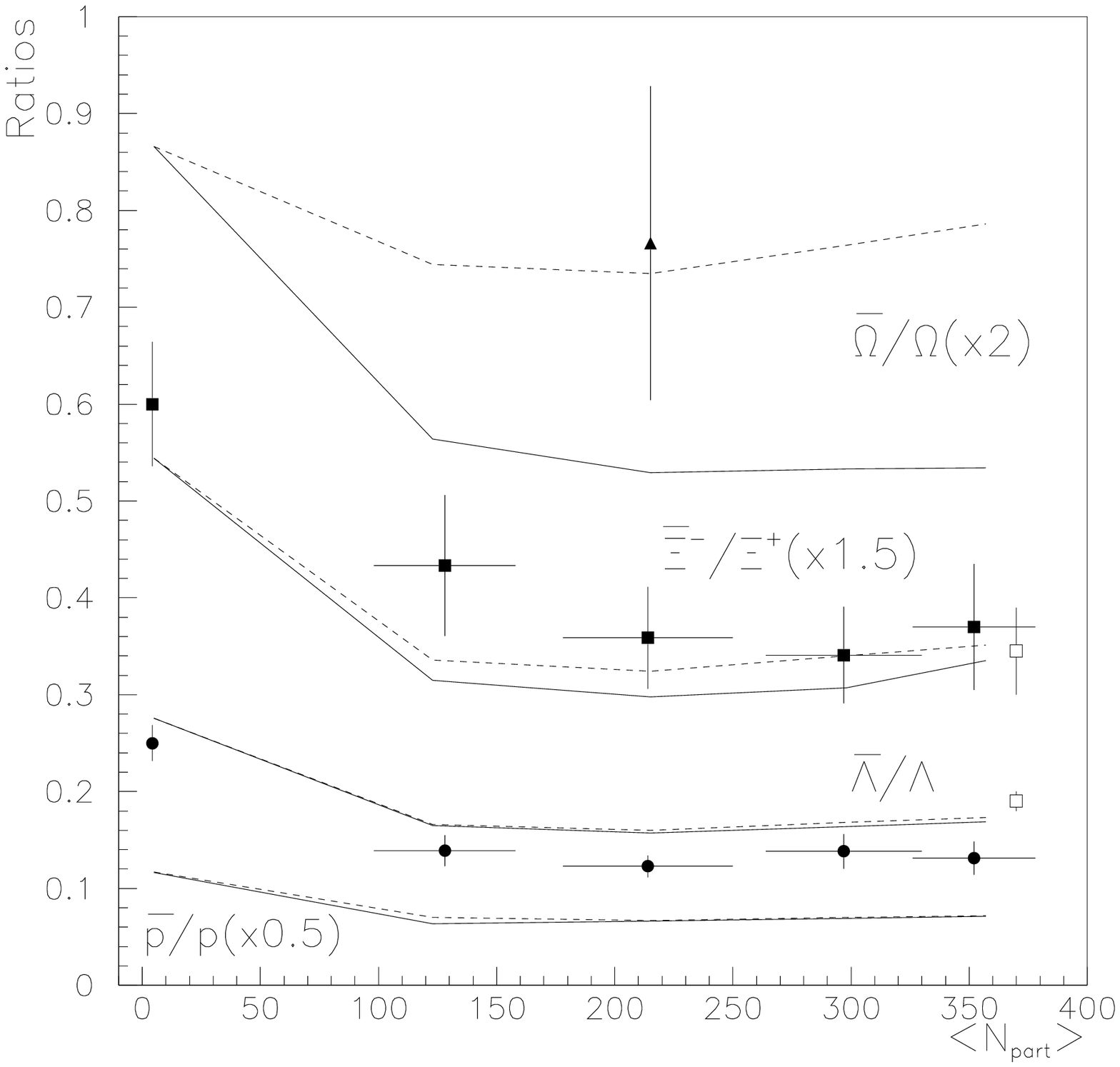,width=13.cm}
\end{center}

\newpage
\centerline{\bf Figure 15}

\begin{center}
\epsfig{file=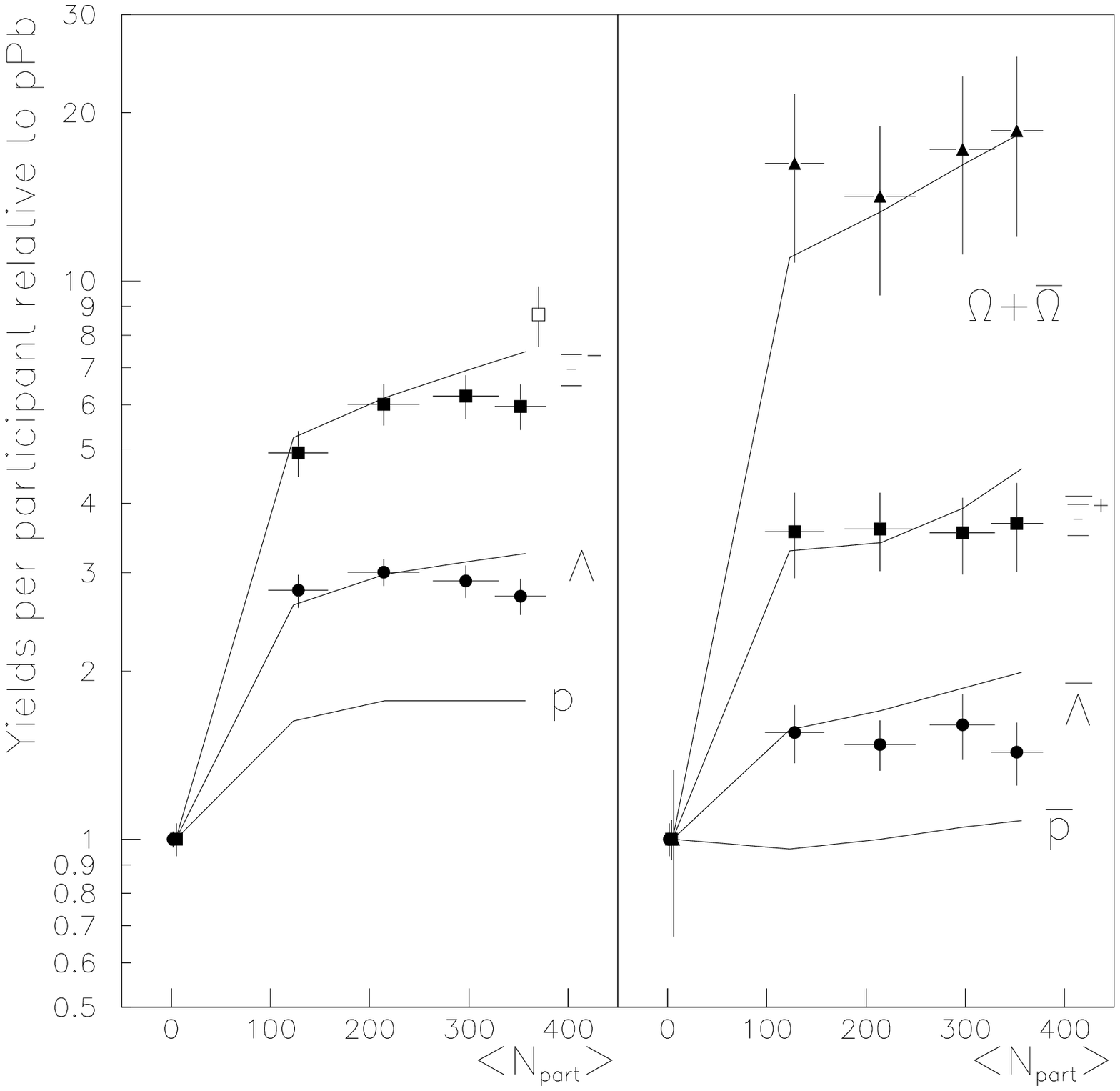,width=13.cm}
\end{center}

\newpage
\centerline{\bf Figure 16a}

\begin{center}
\epsfig{file=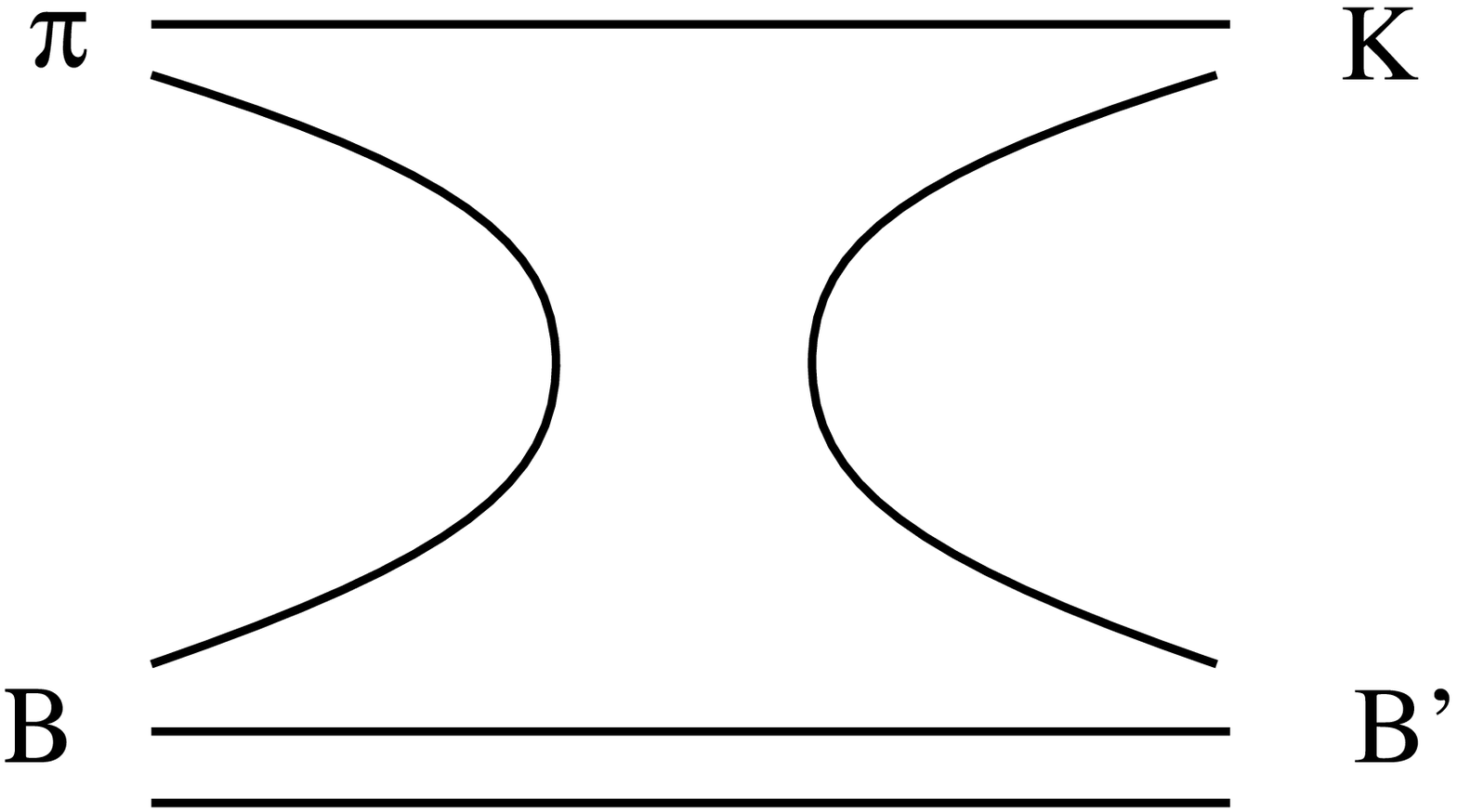,width=7.cm}
\end{center}

\vspace{1cm}
\centerline{\bf Figure 16b}

\begin{center}
\epsfig{file=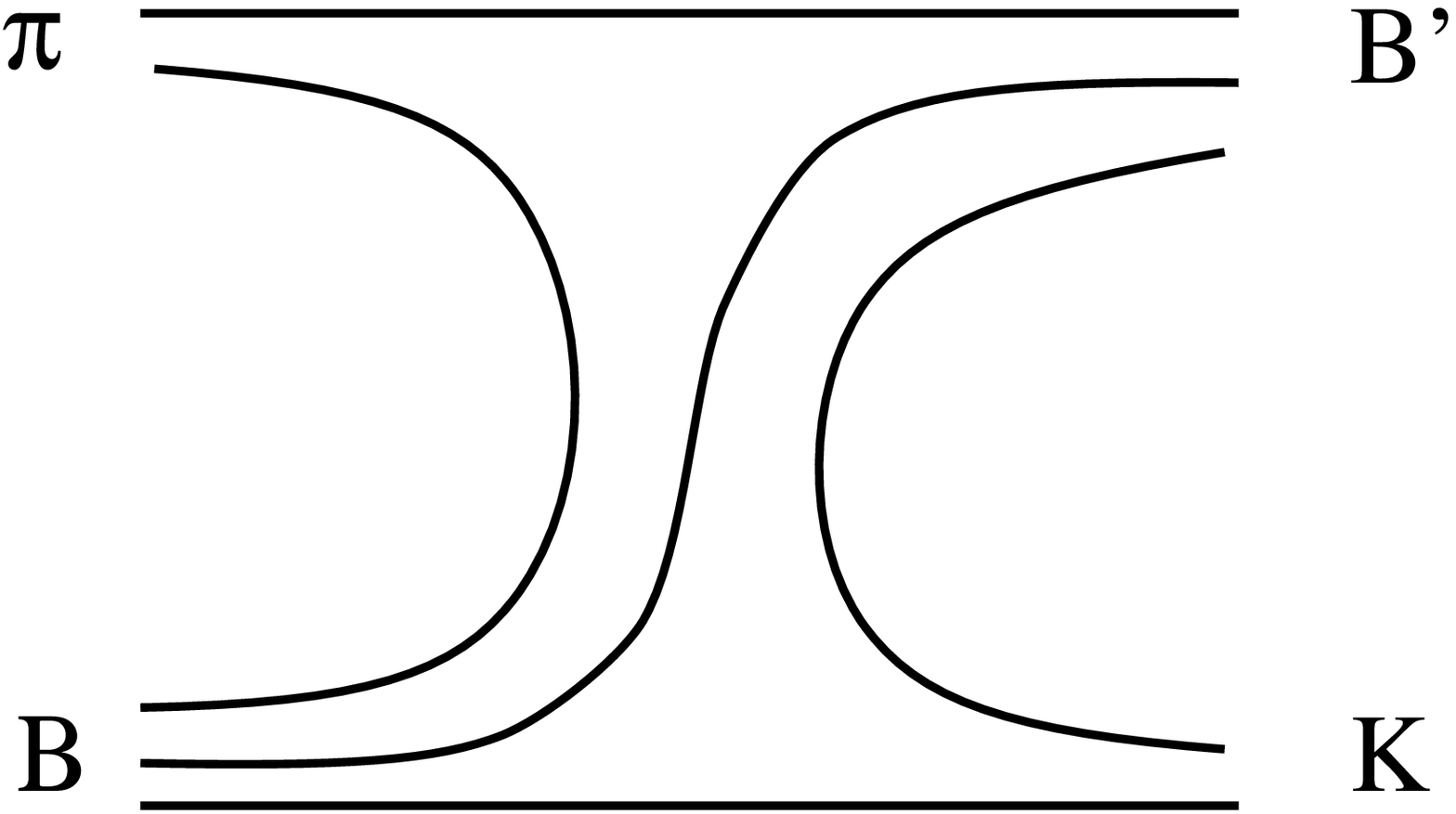,width=7.cm}
\end{center}

\vspace{1cm}
\centerline{\bf Figure 16c}

\begin{center}
\epsfig{file=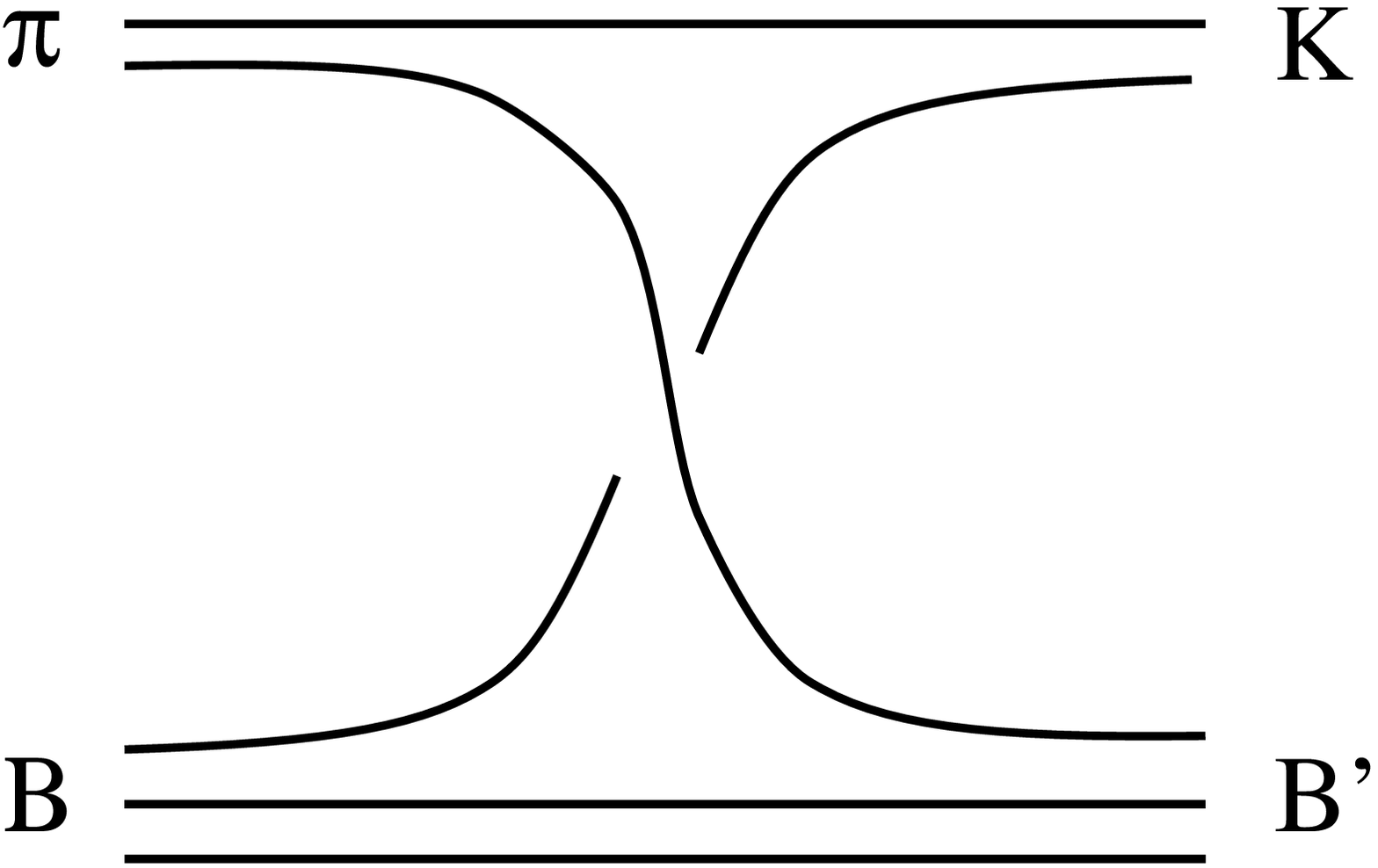,width=7.cm}
\end{center}

\end{document}